\documentclass[epj]{svjour}

\usepackage{graphicx}
\usepackage{fancyhdr}
\def\makeheadbox{{
 }}
\def\mytitle{Colliders and Cosmology} 
\def\myauthors{Keith A. Olive}  
\def\mytype{Plenary}
\def\mysession{Keith A. Olive}

\def\beq{\begin{equation}}
\def\eeq{\end{equation}}
\def\PL{{Phys.~Lett.} }
\def\PR{{Phys.~Rev.} }
\def\PRL{{Phys.~Rev.~Lett.} }

\def\ohsq{\Omega_{\chi} h^2}
\def\m12{m_{1\!/2}}
\def\gev{{\rm \, Ge\kern-0.125em V}}
\def\ga{\mathrel{\raise.3ex\hbox{$>$\kern-.75em\lower1ex\hbox{$\sim$}}}}
\def\la{\mathrel{\raise.3ex\hbox{$<$\kern-.75em\lower1ex\hbox{$\sim$}}}}

\newcommand{\MW}{M_W}

\newcommand{\MA}{M_A}
\newcommand{\Mh}{m_h}
\newcommand{\sweff}{\sin^2\theta_{\mathrm{eff}}}
\def\Ga{\Gamma}
\def\De{\Delta}
\newcommand{\tb}{\tan \beta}
\newcommand{\KL}{\left(}
\newcommand{\KR}{\right)}
\newcommand{\KKL}{\left[}
\newcommand{\KKR}{\right]}
\def\si{\sigma}
\newcommand{\br}{{\rm BR}}
\def\stau{\widetilde \tau}
\def\he#1{\iso{He}{#1}}

\def\li#1{\iso{Li}{#1}}
\def\b1#1{\iso{B}{1#1}}
\newcommand\iso[2]{\mbox{${}^{#2}${\rm #1}}}
\newcommand{\ifb}{\;\mbox{fb}^{-1}}
\newcommand{\TQb}{\tan^2 \beta\hspace{1mm}}
\newcommand{\plane}[2]{($#1, #2$)~plane}

\begin{document}

\title{Colliders and Cosmology}
\author{Keith A. Olive\inst{1}
\thanks{\emph{Email:} olive@umn.edu}%
}                     
%
%
\institute{William I.\ Fine Theoretical Physics Institute,
University of Minnesota, Minneapolis, MN~55455, USA
}
%
\date{}
\abstract{
\vskip - 2in
\rightline{UMN--TH--2617/07}
\rightline{FTPI--MINN--07/27}
\rightline{September 2007}
\vskip 1.5in
Dark matter in variations of constrained 
minimal supersymmetric standard models will be discussed. Particular attention
will be given to the comparison between accelerator and 
direct detection constraints.%
} 
\maketitle
\section{Introduction}
\label{intro}

Evidence for dark matter in the universe is available from a wide range of
observational data.  In addition to the classic evidence from galactic rotation curves \cite{rot},
the analysis of the cosmic microwave background anisotropies leads to the conclusion
that the curvature of the universe is close to zero indicating that the sum of the
fractions of critical density, $\Omega$,  in matter and a cosmological constant (or dark energy)
is close to one \cite{wmap}.  When combined with a variety of data including results from
the analysis of type Ia supernovae observations \cite{sn1} and baryon acoustic oscillations
\cite{ba} one is led to the concordance model where $\Omega_m \sim 0.23$ and $\Omega_\Lambda
\sim 0.73$ with the remainder (leading to $\Omega_{tot} = 1$) in baryonic matter. 
Other dramatic pieces of evidence can be found in combinations of X-ray observations
and weak lensing showing the superposition of dark matter (from lensing) and ordinary matter
from X-ray gas \cite{witt} and from the separation of baryonic and dark matter
after the collision of two galaxies as seen in the Bullet cluster \cite{clowe}.
For a more complete discussion see \cite{otasi3}.

Here, I will adopt the results of
the three-year data from WMAP \cite{wmap} which
has determined many cosmological parameters to
unprecedented precision.  Of particular interest
is the determination of the total matter density (relative to the 
critical density), $\Omega_m h^2$ and the baryonic density, $\Omega_b h^2$.
In the context of the $\Lambda$CDM model, the WMAP only results indicate
\beq
\Omega_m h^2 = 0.1265^{+0.0081}_{-0.0080} \qquad \Omega_b h^2 = 0.0223 \pm 0.0007
\eeq
The difference corresponds to the requisite dark matter density
\beq
\Omega_{CDM} h^2 = 0.1042^{+0.0081}_{-0.0080}
\label{wmap}
\eeq
or a 2$\sigma$ range of 0.0882 -- 0.1204 for $\Omega_{CDM} h^2$.

\section{Constrained MSSM models}
\label{sec:1} 
In its generality, the minimal supersymmetric standard model (MSSM) has over 100 undetermined
parameters. But in addition to relieving the helplessness of an analysis with so many
free parameters, there are good arguments based on grand unification \cite{gut} and supergravity \cite{BIM} which lead to a strong reduction in the number of parameters.
I will assume several unification conditions placed on the
supersymmetric parameters.  In all models considered, the gaugino masses
are assumed to be unified at the GUT scale with value, $m_{1/2}$, 
as are the trilinear couplings with value $A_0$.  Also common to all models
considered here is the unification of all soft scalar masses set equal to $m_0$ at the GUT scale.
With this set of boundary conditions at the GUT scale, we can use the the radiative electroweak 
symmetry breaking conditions by specifying the ratio of the two Higgs 
vacuum expectation values, $\tan \beta$, and the mass, $M_Z$, to predict the 
values of the Higgs mixing mass parameter, $\mu$ and the bilinear coupling, $B$.
The sign of $\mu$ remains free. 
This class of models is often referred to as the constrained MSSM (CMSSM) \cite{funnel,cmssm,efgosi,eoss,cmssmwmap}.

Although the CMSSM is often confused with in name with mSUGRA, i.e. models based on minimal
supergravity, \cite{bfs}, the latter employ two additional constraints \cite{vcmssm}.
In addition to the conditions listed above,
these are: a relation between the bilinear and trilinear couplings $B_0 = A_0 - m_0$, and the
relation between the gravitino mass and soft scalar masses, $m_{3/2} = m_0$.
When electroweak symmetry breaking boundary conditions are applied,
this theory contains only $m_{1/2}, m_0$, and $A_0$ in addition to the sign of 
$\mu$, as free parameters. The magnitude of $\mu$ as well
as $\tan \beta$ are predicted.   

 I will also discuss a less constrained model, the NUHM, in which the Higgs soft
masses are not unified at the GUT scale \cite{nonu,nuhm}.  In this class of models,
both $\mu$ and the Higgs pseudo scalar mass become free parameters. 

In all of the models discussed below, I will assume unbroken $R-parity$ and
and hence the lightest supersymmetric particle is stable. This will often, but not
always be the neutralino \cite{EHNOS}. 

\section{The CMSSM after WMAP}
\label{sec:2} 

\begin{figure}[h]
\includegraphics[height=3in]{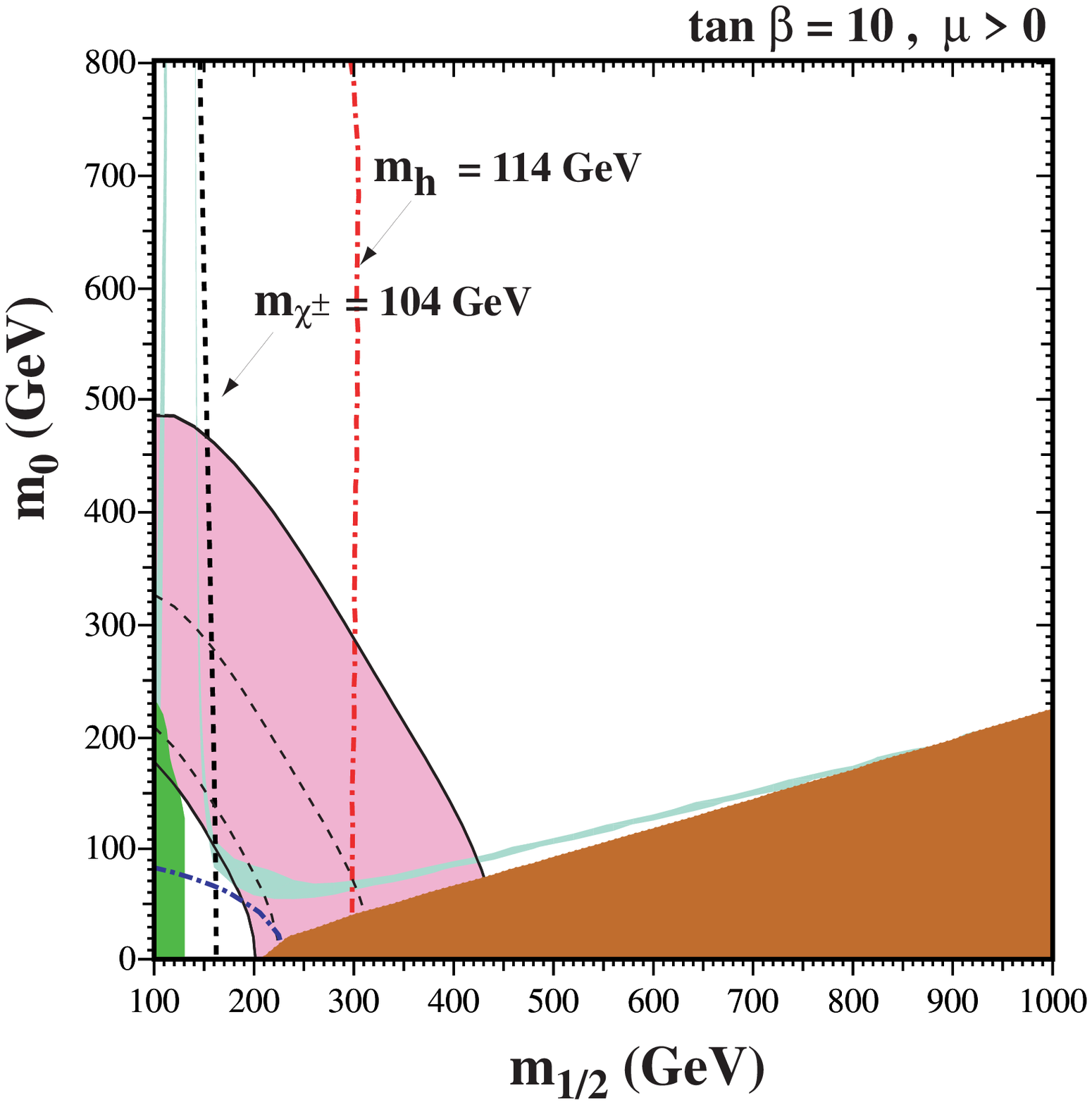}
\includegraphics[height=3in]{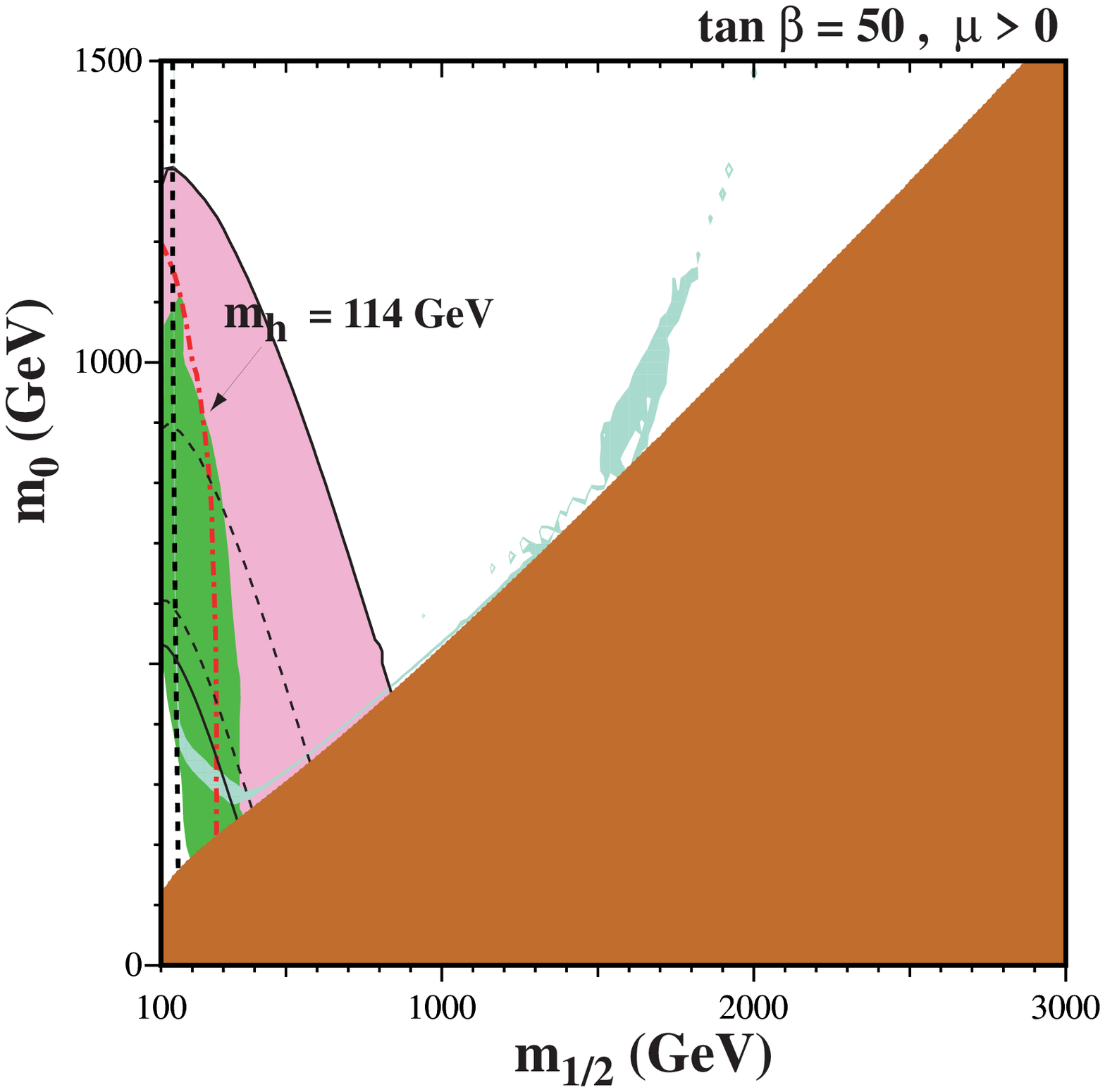}
\caption{\label{fig:UHM}
{\it The $(m_{1/2}, m_0)$ planes for  (a) $\tan \beta = 10$ and  $\mu > 0$, 
assuming $A_0 = 0, m_t = 175$~GeV and
$m_b(m_b)^{\overline {MS}}_{SM} = 4.25$~GeV. The near-vertical (red)
dot-dashed lines are the contours $m_h = 114$~GeV, and the near-vertical (black) dashed
line is the contour $m_{\chi^\pm} = 104$~GeV. Also
shown by the dot-dashed curve in the lower left is the corner
excluded by the LEP bound of $m_{\tilde e} > 99$ GeV. The medium (dark
green) shaded region is excluded by $b \to s
\gamma$, and the light (turquoise) shaded area is the cosmologically
preferred region. In the dark
(brick red) shaded region, the LSP is the charged ${\tilde \tau}_1$. The
region allowed by the E821 measurement of $a_\mu$ at the 2-$\sigma$
level, is shaded (pink) and bounded by solid black lines, with dashed
lines indicating the 1-$\sigma$ ranges. In (b), $\tan \beta= 50$. }}
\end{figure}

For a given value of $\tan \beta$, $A_0$,  and $sgn(\mu)$, the resulting regions of 
acceptable relic density and which satisfy the phenomenological constraints
can be displayed on the  $m_{1/2} - m_0$ plane.
In Fig. \ref{fig:UHM}a,  the light
shaded region corresponds to that portion of the CMSSM plane
with $\tan \beta = 10$, $A_0 = 0$, and $\mu > 0$ such that the computed
relic density yields the WMAP value given in eq. (\ref{wmap}) \cite{eoss}.
The bulk region at relatively low values of 
$m_{1/2}$ and $m_0$,  tapers off
as $\m12$ is increased.  At higher values of $m_0$,  annihilation cross sections
are too small to maintain an acceptable relic density and $\ohsq$ is too large.
Although sfermion masses are also enhanced at large $\m12$ (due to RGE running),
co-annihilation processes between the LSP and the next lightest sparticle 
(in this case the $\tilde \tau$) enhance the annihilation cross section and reduce the
relic density.  This occurs when the LSP and NLSP are nearly degenerate in mass.
The dark shaded region has $m_{\tilde \tau}< m_\chi$
and is excluded.   The effect of coannihilations is
to create an allowed band about 25-50 $\gev$ wide in $m_0$ for $\m12 \la
950\gev$, or $\m12 \la 400\gev$, which tracks above the $m_{{\tilde \tau}_1}
 =m_\chi$ contour~\cite{efo}.

Also shown in Fig. \ref{fig:UHM}a are
the relevant phenomenological constraints.  
These include the LEP limits on the chargino mass: $m_{\chi^\pm} > 104$~GeV~\cite{LEPsusy}, 
on the selectron mass: $m_{\tilde e} > 99$~GeV~ \cite{LEPSUSYWG_0101} 
and on the Higgs mass: $m_h >
114$~GeV~\cite{LEPHiggs}. The former two constrain $m_{1/2}$ and $m_0$ directly
via the sparticle masses, and the latter indirectly via the sensitivity of
radiative corrections to the Higgs mass to the sparticle masses,
principally $m_{\tilde t, \tilde b}$. 
{\tt FeynHiggs}~\cite{FeynHiggs} is used for the calculation of $m_h$. 
The Higgs limit  imposes important constraints
principally on $m_{1/2}$ particularly at low $\tan \beta$.
Another constraint is the requirement that
the branching ratio for $b \rightarrow
s \gamma$ is consistent with the experimental measurements \cite{bsgex}. 
These measurements agree with the Standard Model, and
therefore provide bounds on MSSM particles \cite{gam},  such as the chargino and
charged Higgs masses, in particular. Typically, the $b\rightarrow s\gamma$
constraint is more important for $\mu < 0$, but it is also relevant for
$\mu > 0$,  particularly when $\tan\beta$ is large. The constraint imposed by
measurements of $b\rightarrow s\gamma$ also excludes small
values of $m_{1/2}$. 
Finally, there are
regions of the $(m_{1/2}, m_0)$ plane that are favoured by
the BNL measurement \cite{newBNL} of $g_\mu - 2$ at the 2-$\sigma$ level, corresponding to 
a deviation  from the Standard Model 
calculation~\cite{Davier} using $e^+ e^-$ data.

Another
mechanism for extending the allowed regions in the CMSSM to large
$m_\chi$ is rapid annihilation via a direct-channel pole when $m_\chi
\sim {1\over 2} m_{A}$~\cite{funnel,efgosi}. Since the heavy scalar and
pseudoscalar Higgs masses decrease as  
$\tan \beta$ increases, eventually  $ 2 m_\chi \simeq  m_A$ yielding a
`funnel' extending to large
$m_{1/2}$ and
$m_0$ at large
$\tan\beta$, as seen in Fig.~\ref{fig:UHM}b.
As one can see, the impact of the Higgs mass constraint is reduced (relative to 
the case with $\tan \beta = 10$) while that of $b \to s \gamma$ is enhanced.

\begin{figure}
\begin{center}
\includegraphics[height=3in]{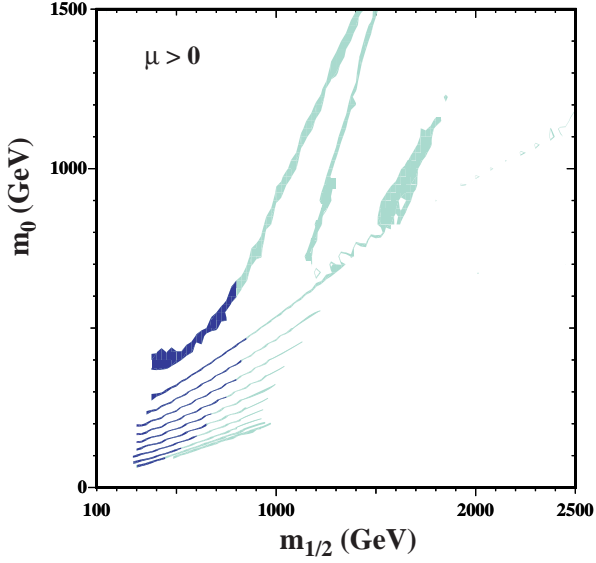}
\end{center}
\caption{\label{fig:strips}\it
The strips display the regions of the $(m_{1/2}, m_0)$ plane that are
compatible with the WMAP determination of $\ohsq$ and the laboratory
constraints for $\mu > 0$ and $\tan \beta = 5, 10, 15, 20, 25, 30,
35, 40, 45, 50, 55$. The parts of the strips compatible with $g_\mu - 2$ 
at the 2-$\sigma$ level have darker shading.
}
\end{figure}

Shown in Fig.~\ref{fig:strips} are the WMAP lines \cite{eoss} of the $(m_{1/2}, m_0)$
plane for $\mu > 0$ and
values of $\tan \beta$ from 5 to 55, in steps $\Delta ( \tan \beta ) = 5$.
We notice immediately that the strips are considerably narrower than the
spacing between them, though any intermediate point in the $(m_{1/2},
m_0)$ plane would be compatible with some intermediate value of $\tan
\beta$. The right (left) ends of the strips correspond to the maximal
(minimal) allowed values of $m_{1/2}$ and hence $m_\chi$. 
The lower bounds on $m_{1/2}$ are due to the Higgs 
mass constraint for $\tan \beta \le 23$, but are determined by the $b \to 
s \gamma$ constraint for higher values of $\tan \beta$.

\begin{figure}
\begin{center}
\includegraphics[height=3in]{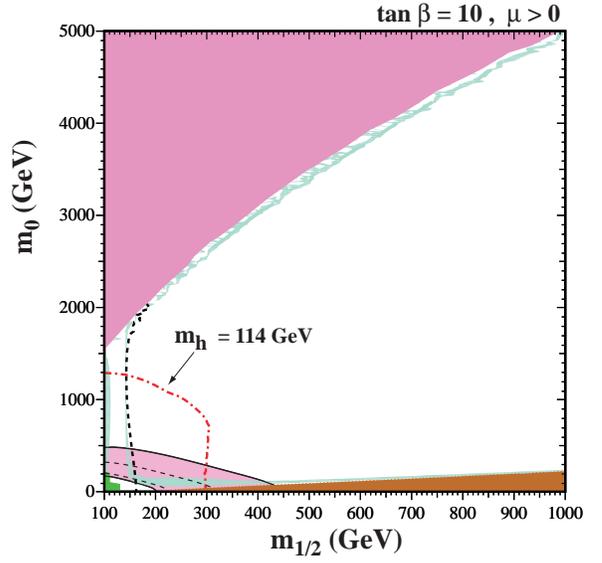}
\end{center}
\caption{\label{figfp}\it
As in Fig. \protect\ref{fig:UHM}a, where the range in $m_0$ is extended to
5 TeV.  In the shaded region at very high $m_0$, there are no solutions 
for $\mu$ which respect the low energy electroweak symmetry breaking conditions.
}
\end{figure}

Finally, there is one additional region of acceptable relic density known as the
focus-point region \cite{fp}, which is found
at very high values of $m_0$. An example showing this region is found in Fig. \ref{figfp},
plotted for $\tan \beta = 10$, $\mu > 0$, and $m_t = 175$ TeV.
As $m_0$ is increased, the solution for $\mu$ at low energies as determined
by the electroweak symmetry breaking conditions eventually begins to drop. 
When $\mu \la m_{1/2}$, the composition of the LSP gains a strong Higgsino
component and as such the relic density begins to drop precipitously. 
As $m_0$ is increased further, there are no longer any solutions for $\mu$.  This 
occurs in the shaded region in the upper left corner of Fig. \ref{figfp}. 

\section{Direct detection}

Direct detection techniques
rely on the neutralino nucleon scattering cross-section.
In general, there are two contributions to the low-energy effective
four-fermion
Lagrangian which are not velocity dependent. 
These can be expressed as spin-dependent and scalar or spin-independent interactions.
Here I will discuss only spin-independent elastic $\chi$-nucleon   
scattering stemming from
\begin{equation}
{\cal L} \, = \, \alpha_{3i} \bar{\chi} \chi \bar{q_{i}} q_{i},
\label{lagr}
\end{equation}
which is to be summed over the quark flavours $q$, and the
subscript $i$ labels up-type quarks ($i=1$) and down-type quarks
($i=2$). The expression for $\alpha_{3i}$ can be found in \cite{eflo}.

The scalar cross section can be written in terms of $\alpha_{3i}$,
\begin{equation}
\sigma_{3} = \frac{4 m_{r}^{2}}{\pi} \left[ Z f_{p} + (A-Z) f_{n}
\right]^{2} ,
\label{si}
\end{equation}
where $m_r$ is the reduced LSP mass,
\begin{equation}
\frac{f_{p}}{m_{p}} = \sum_{q=u,d,s} f_{Tq}^{(p)}
\frac{\alpha_{3q}}{m_{q}} +
\frac{2}{27} f_{TG}^{(p)} \sum_{c,b,t} \frac{\alpha_{3q}}{m_q},
\label{f}
\end{equation}
the parameters $f_{Tq}^{(p)}$  are defined by
\begin{equation}
m_p f_{Tq}^{(p)} \equiv \langle p | m_{q} \bar{q} q | p \rangle
\equiv m_q B_q ,
\label{defbq}
\eeq
$f_{TG}^{(p)} = 1 - \sum_{q=u,d,s} f_{Tq}^{(p)} $~\cite{SVZ},
and $f_{n}$ has a similar expression.  
The needed matrix elements are determined in part by 
the $\pi$-nucleon $\Sigma$ term, which is 
given by
\beq
\sigma_{\pi N} \equiv \Sigma = {1 \over 2} (m_u + m_d) (B_u + B_d) .
\eeq
and the strangeness contribution to the proton mass, 
\beq
y = {2 B_s \over B_u + B_d} = 1 - {\sigma_0 \over \Sigma}
\eeq
where 
$\sigma_0$ is the change in the nucleon mass due to the 
non-zero $u, d$ quark masses, which is estimated on the basis of octet 
baryon mass differences to be $\sigma_0 = 36 
\pm 7$ MeV~\cite{oldsnp}.
Values of the matrix elements are given in the table for 3 choices of $\Sigma$ and hence $y$.
As one can see, we expect there will be a strong dependence of the cross section 
on the proton strangeness, $y$.

\begin{table}
\caption{Matrix elements in terms of proton strangeness}
\label{tab:1}       
\begin{tabular}{lllll}
\hline\noalign{\smallskip}
$\Sigma$ (MeV) & $y$ & $f_{T_u}$ & $f_{T_d}$ & $f_{T_s}$   \\
\noalign{\smallskip}\hline\noalign{\smallskip}
45 & 0.2 & 0.020 & 0.026 & 0.117 \\
64 &0.44 & 0.027 & 0.039 & 0.363 \\
36 & 0 & 0.016 & 0.020 & 0.0 \\
\noalign{\smallskip}\hline
\end{tabular}
\end{table}

\section{Indirect sensitivities}

Measurements of electroweak precision observables~\, (EWPO) as well as 
B-physics observables (BPO) can provide interesting indirect information
about the supersymmetric parameter space. We have already seen the impact of
measurements of the anomolous magnetic moment of the muon, the branching raio of
$b \to s \gamma$ in addition to the non-discovery of charginos and the Higgs boson at 
LEP which  impose significant lower bounds on $m_{1/2}$. 

In \cite{ehow3,ehow4,ehoww}, 
we considered the following EWPO: the $W$~boson mass, $\MW$, the
effective weak mixing angle at the $Z$~boson resonance, $\sweff$, 
the width of the $Z$, $\Gamma_Z$, the Higgs mass, $m_h$, and 
 \mbox{$(g_\mu-2)$} in addtion to the BPO: 
$b$ decays $b \to s \gamma$, $B_s \to \mu^+\mu^-$, 
$B_u \to \tau \nu_\tau$ and the $B_s$ mass mixing parameter $\De M_{B_s}$. 
We performed the analysis of the sensitivity to $m_{1/2}$ moving 
along the WMAP
strips with fixed values of $A_0$ and $\tb$. The experimental central
values, the present experimental errors and theoretical uncertainties
are as described in \cite{ehoww}.
Assuming that the nine observables listed above are
uncorrelated, a $\chi^2$ fit 
has been performed with
\begin{eqnarray}
\chi^2 & \equiv & \sum_{n=1}^{7} \KKL \KL
              \frac{R_n^{\rm exp} - R_n^{\rm theo}}{\si_n} \KR^2
              + 2 \log \KL \frac{\si_n}{\si_n^{\rm min}} \KR \KKR \nonumber \\
                                           &&    + \chi^2_{\Mh}
                                               + \chi^2_{B_s}.
\label{eq:chi2}
\end{eqnarray}
Here $R_n^{\rm exp}$ denotes the experimental central value of the
$n$th observable ($\MW$, $\sweff$, $\Ga_Z$, \mbox{$(g-2)_\mu$} and
$\br(b \to s \gamma)$, $\br(B_u \to \tau \nu_\tau$), $\De M_{B_s}$),
$R_n^{\rm theo}$ is the corresponding MSSM prediction and $\si_n$
denotes the combined error. 
Additionally,
$\si_n^{\rm min}$ is the minimum combined error over the parameter space of
each data set, and
$\chi^2_{\Mh}$ and $\chi^2_{B_s}$ denote the $\chi^2$ contribution
coming from the experimental limits 
on the lightest MSSM Higgs boson mass and on $\br(B_s \to \mu^+\mu^-)$,
respectively \cite{ehoww}.

Details of the analysis \cite{ehoww} were discussed in \cite{sven,georg}
and here I show the net result of the $\chi^2$ analysis in the CMSSM
in Fig. \ref{fig:chi} using the combined $\chi^2$~values
for the EWPO and BPO, computed from Eq. \ref{eq:chi2}.
We see that the global minimum of $\chi^2 \sim 4.5$
for both values of $\tb$. This is quite a good fit for the number of
experimental observables being fitted.
For both values of $\tb$, the focus-point region is disfavoured by comparison
with the coannihilation region, though this effect is less important for
$\tb = 50$. 
For $\tb = 10$, $m_{1/2} \sim 300 \gev$ and $A_0 > 0$ are preferred, whereas,
for $\tb = 50$, $m_{1/2} \sim 600 \gev$ and $A_0 < 0$ are preferred. This
change-over is largely due to the impact of the LEP $\Mh$ constraint for
$\tb = 10$ and the $b \to s \gamma$ constraint for $\tb = 50$.

\begin{figure}[tbh!]
\begin{center}
\includegraphics[height=2.8in]{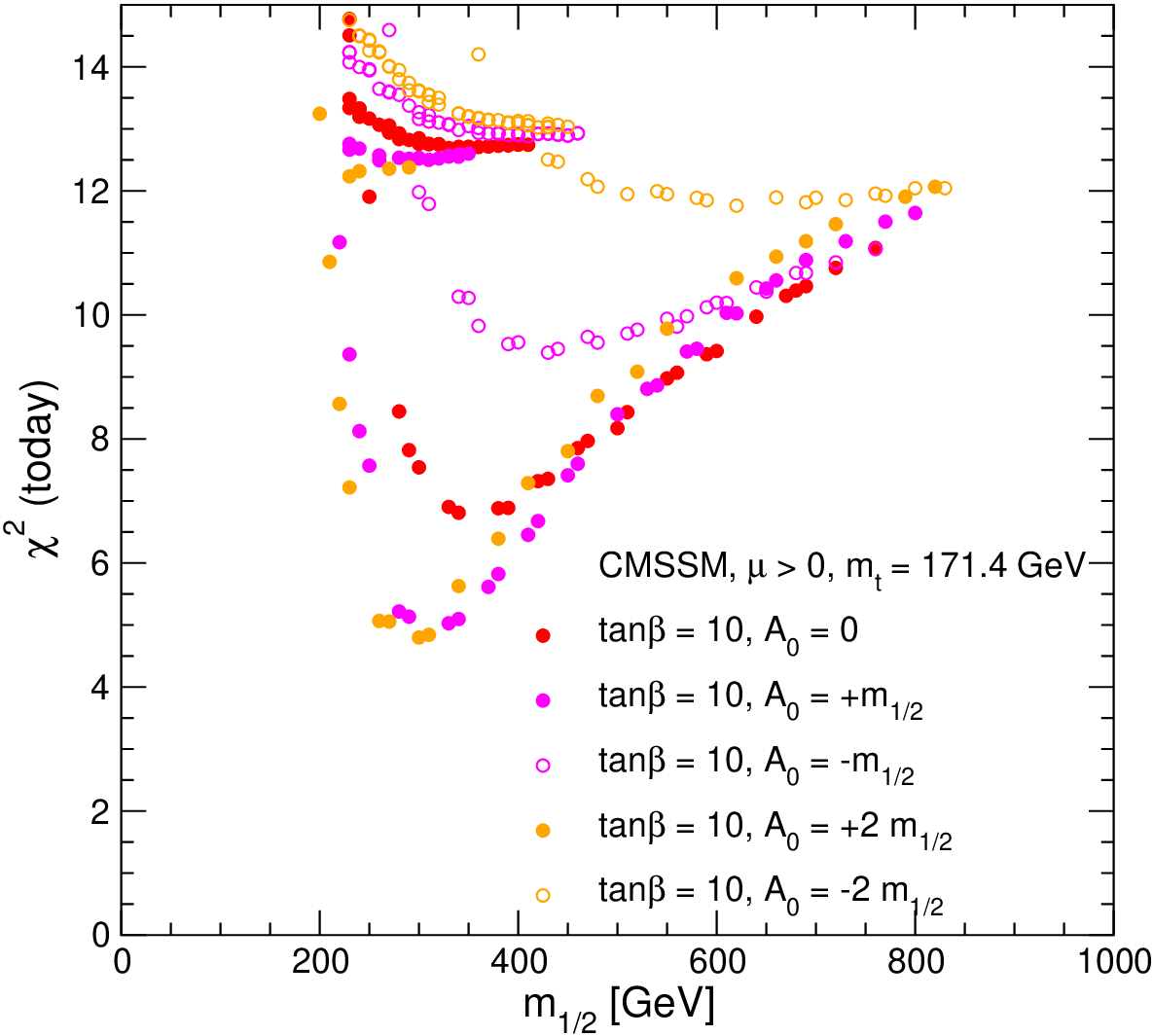}
\includegraphics[height=2.8in]{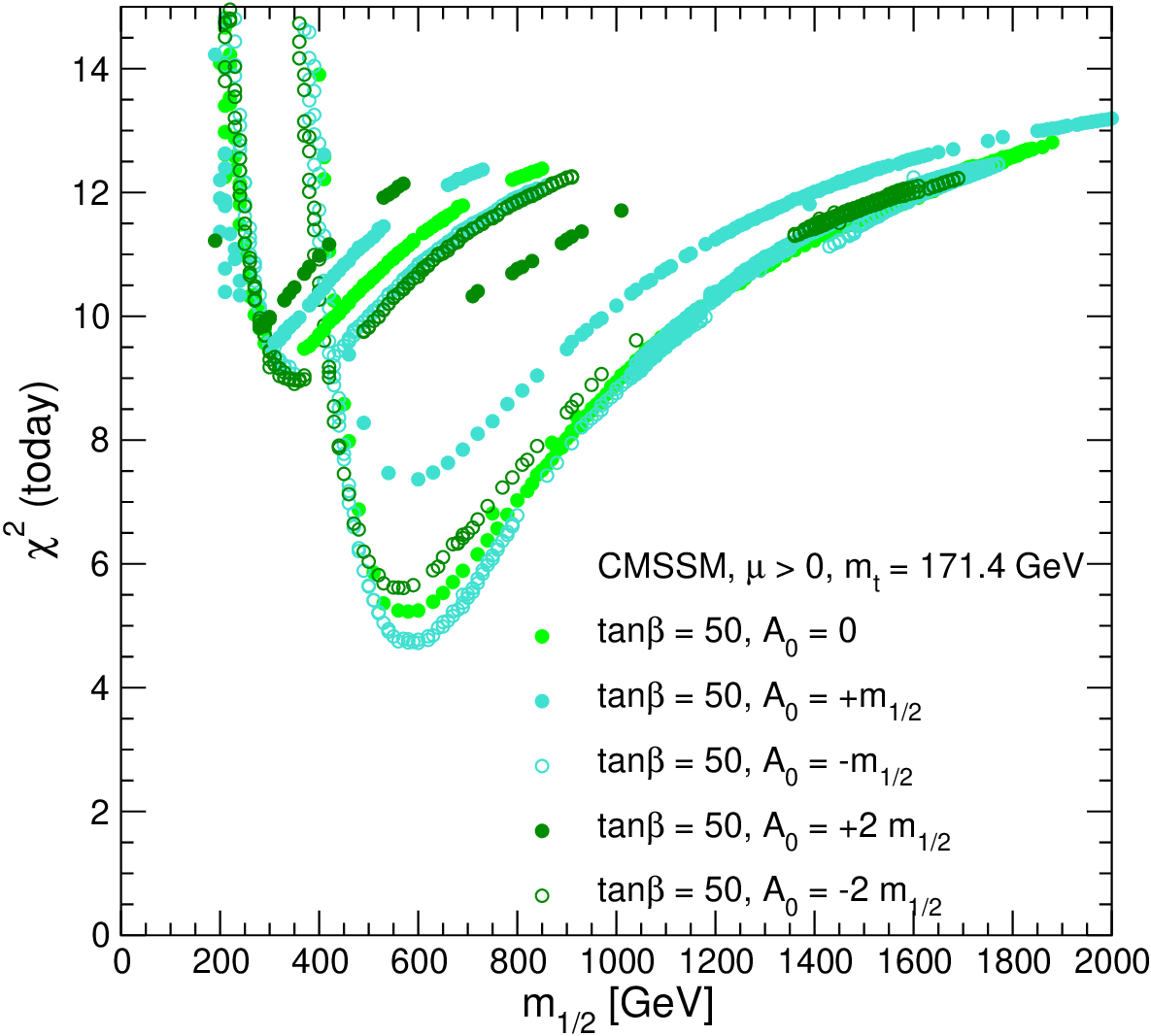}
\caption{%
The combined $\chi^2$~function for the electroweak
observables $\MW$, $\sweff$, $\Ga_Z$, $(g - 2)_\mu$, $\Mh$, 
and the $b$~physics observables
$\br(b \to s \gamma)$, $\br(B_s \to \mu^+\mu^-)$, $\br(B_u \to \tau \nu_\tau)$
and $\De M_{B_s}$, evaluated in the CMSSM for $\tb = 10$ (a) and
$\tb = 50$ (b) for various discrete values of $A_0$. We use
$m_t = 171.4 \pm 2.1 $ GeV and $m_b(m_b) = 4.25 \pm 0.11$ GeV, and 
$m_0$ is chosen to yield the central value of the cold dark matter
density indicated by WMAP and other observations for the central values
of $m_t$ and $m_b(m_b)$.}
\label{fig:chi}
\end{center}
\vspace{-1em}
\end{figure}

\section{Direct Detection in the CMSSM}

In Fig.~\ref{fig:Andyall}, we display the expected ranges of 
the spin-independent  cross sections in the CMSSM when we
sample randomly $\tan \beta$ as well as the other CMSSM parameters \cite{eoss8}. 
All points shown satisfy all phenomenological constraints and have $\Omega_\chi h^2$ 
less than the WMAP upper limit.  Since models with low $\Omega_\chi h^2$ can not be excluded
if another form of dark matter is assumed to exist, these models have been included,
but their cross sections have been scaled by a factor $\Omega_\chi/\Omega_{CDM}$.
In  Fig.~\ref{fig:Andyall}a, $\Sigma_{\pi N} = 45$ MeV has been chosen, and in  \ref{fig:Andyall}b
results for $\Sigma_{\pi N} = 64$ MeV are shown for comparison. As one can see,
the cross sections shift higher by a factor of a few for the larger value of $\Sigma_{\pi N} $.

Also shown on the plot are the current CDMS \cite{cdms} and XENON10 \cite{xenon}
exclusion curves
which place an upper limit on the scattering cross section.  As one can 
see, the current limits have only just now begun to probe
CMSSM models.

\begin{figure}[h]
\centering
\includegraphics[height=3in]{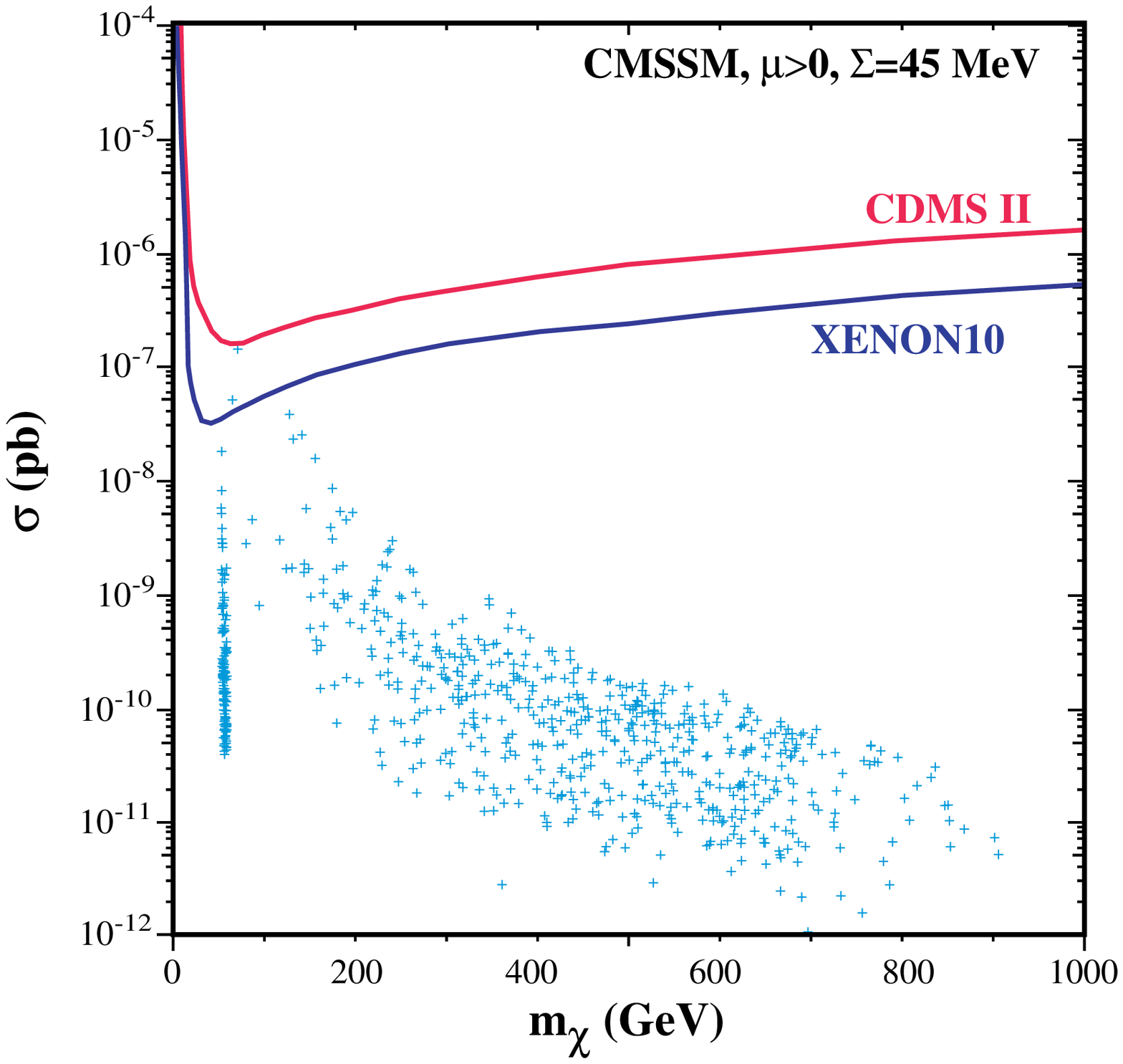}
\includegraphics[height=3in]{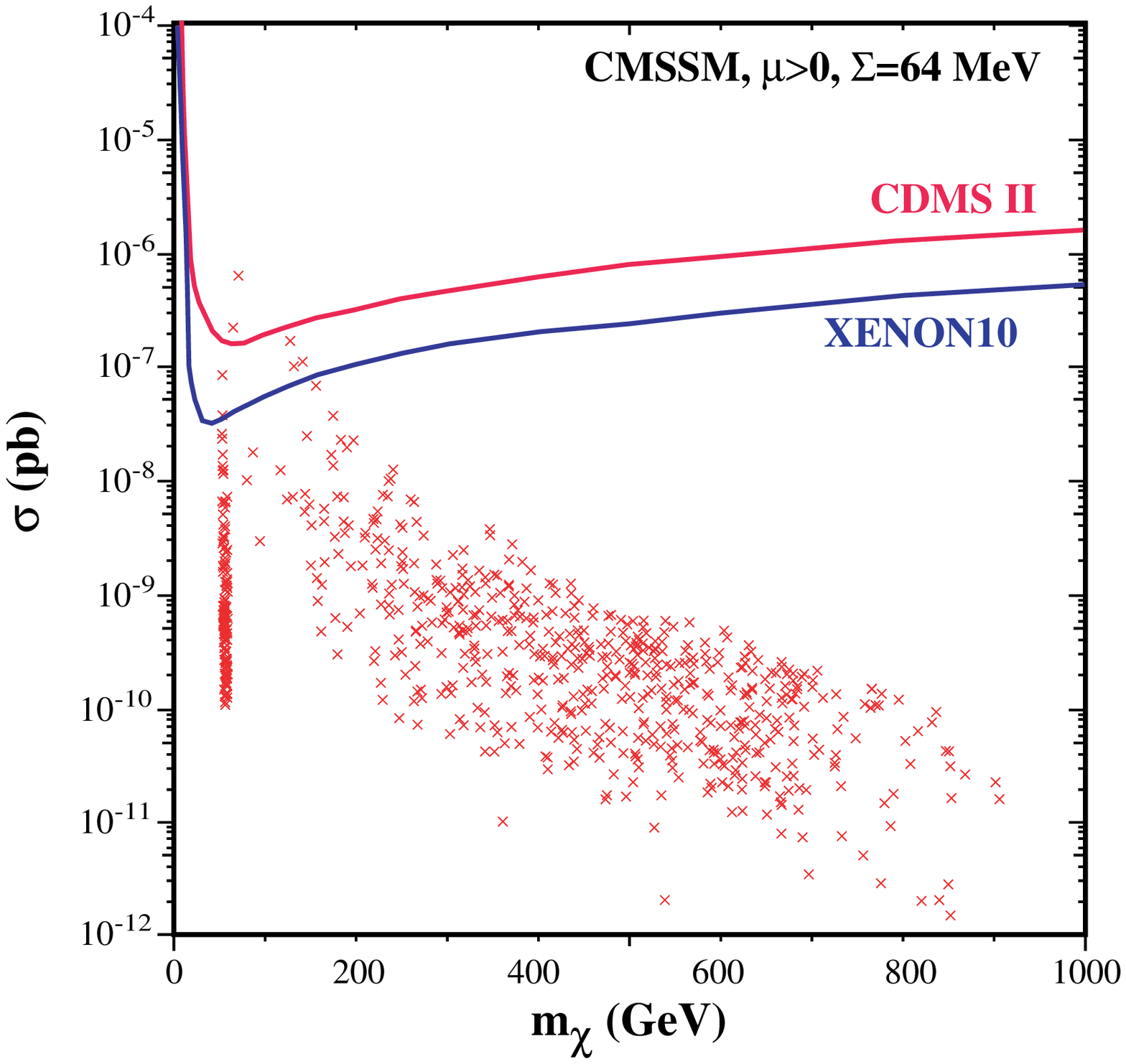}
\caption{\label{fig:Andyall}
{\it 
Scatter plots of the spin-independent elastic-scattering cross
section predicted in the CMSSM for (a) $\Sigma_{\pi N} = 45$ MeV and (b) 64 MeV.}}
\end{figure}

\section{GUT-less models}

The input scale at which universality is assumed in CMSSM models is usually taken to be the
SUSY GUT scale, $M_{GUT} \sim 2 \times 10^{16}$ GeV. However, it may be more
appropriate in some models to assume the soft SUSY-breaking parameters to be
universal at some different input scale, $M_{in}$.
Specific scenarios in which the soft SUSY-breaking parameters may be
universal at a scale below $M_{GUT}$ occur in models with mixed
modulus-anomaly mediated SUSY breaking, also called mirage-mediation~\cite{mixed},
and models with warped extra dimensions~\cite{itoh}.  In the case of mirage-mediation, the
universality scale is the mirage messenger scale, which is predicted
to be $M_{in} \sim 10^{10}-10^{12}$ GeV in the case of KKLT-style moduli
stabilization~\cite{KKLT}. In other models, the universality scale may lie
anywhere between 1 TeV and $M_{Pl}$.

In the CMSSM with universality
imposed at the GUT scale, the one-loop renormalizations of the
gaugino masses $M_a$, where $a=1,2,3$, are the same as those for the
corresponding gauge couplings, $\alpha_a$. Thus, 
at the one-loop level the gaugino masses at
any scale $Q \leq M_{GUT}$ can be expressed as 
\beq
M_a(Q) = \frac{\alpha_a(Q)}{\alpha_a(M_{GUT})}M_a(M_{GUT}),
\eeq
where $M_a(M_{GUT}) = m_{1/2}$. On the other hand, in a
GUT-less CMSSM, where the gauge-coupling strengths run at all scales
below the GUT scale but the soft SUSY-breaking parameters run only below
the lower universality scale, $M_{in}$, at which all the gaugino masses are
assumed to be equal to $m_{1/2} = M_a(M_{in})$, we have
\beq
M_a(Q) = \frac{\alpha_a(Q)}{\alpha_a(M_{in})}m_{1/2}
\label{gaugino}
\eeq
at the one-loop level.
Since the runnings of the coupling strengths in GUT and GUT-less CMSSM
scenarios are identical, the low-energy effective soft gaugino
masses, $M_a(Q)$, in GUT-less cases are less separated and closer to
$m_{1/2}$ than in the usual GUT CMSSM. 
Similarly, the sfermion masses also run less and have masses closer to $m_0$ at
the weak scale than their corresponding masses in the CMSSM.

One of the most
dramatic changes when $M_{in}$ is lowered from the GUT scale is 
its effect on the footprint in the $(m_{1/2}, m_0)$ plane
of the constraint on the relic abundance of
neutralinos inferred from WMAP.
In the GUT-less CMSSM scenario \cite{eosk}, 
as the universality scale is lowered to $M_{in} \sim 10^{12}$
GeV, the co-annihilation strip, the funnel and the focus point regions of the CMSSM 
approach each other and merge, forming a small WMAP-preferred
island in a sea of parameter space where the neutralino relic density
is too small to provide all the cold dark matter wanted by WMAP.

\begin{figure}[ht!]
\includegraphics[height=3in]{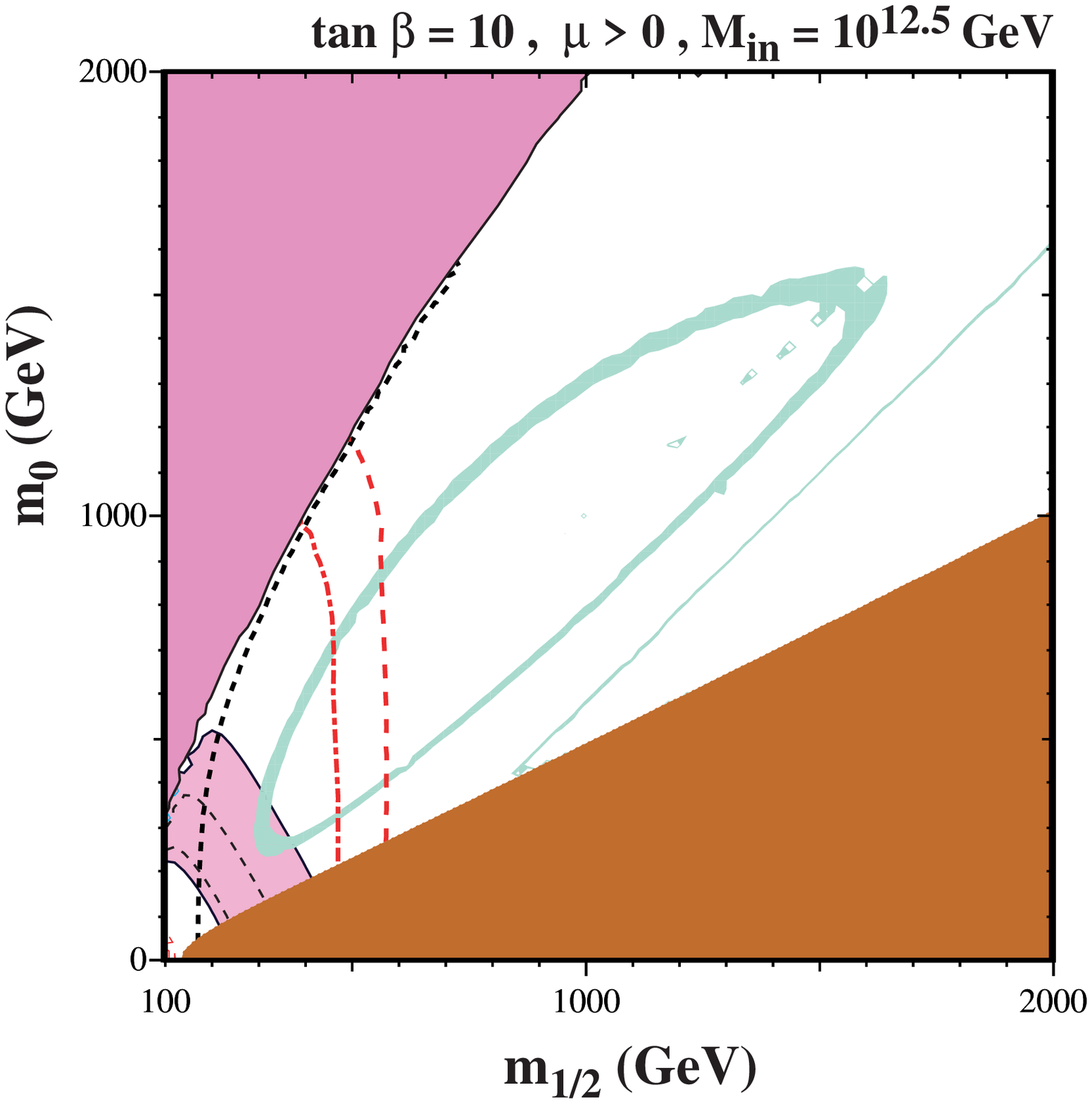}
\includegraphics[height=3in]{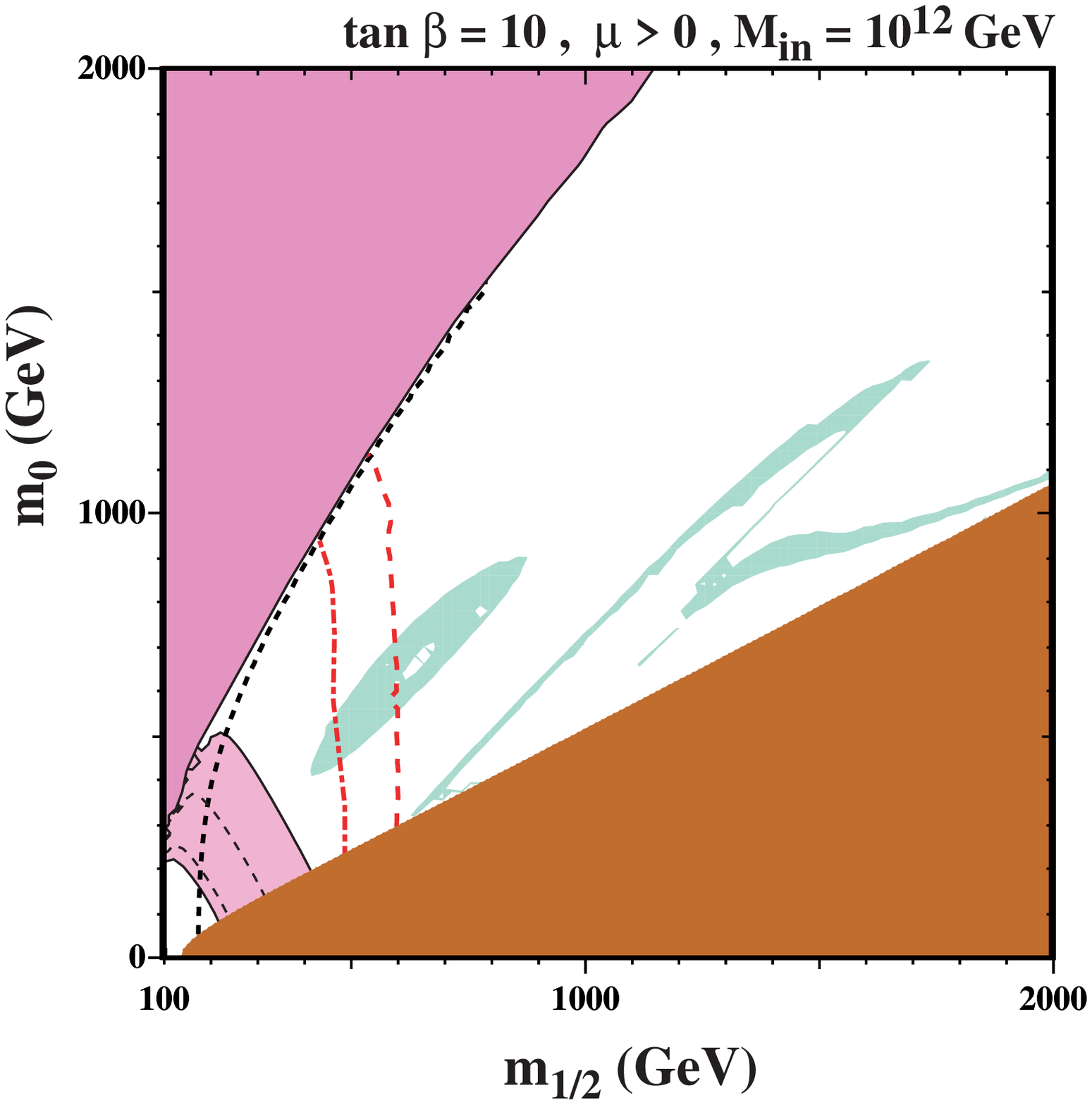}
\caption{\label{fig:mint}
{\it 
Examples of $(m_{1/2}, m_0)$ planes with $\tan \beta = 10$ and 
$A_0 = 0$ but with  different values of $M_{in}$.
(a)  $M_{in} = 10^{12.5}$ GeV, 
(b) $M_{in} = 10^{12}$ GeV. 
In each panel, we show contours representing the LEP lower limits on the
chargino mass (black dashed line), a Higgs mass of 114 GeV (red dashed),
and the more exact (and relaxed) Higgs bound (red dot-dashed). We also show the region
ruled out because the LSP would be charged (dark red shading), and
that excluded by the electroweak vacuum condition (dark pink shading). The region favoured 
by WMAP has light turquoise shading, and the region 
suggested by $g_\mu - 2$ at 2-$\sigma$ has medium (pink) shading, with
the 1-$\sigma$ contours shown as black dashed lines. }}
\end{figure}

As found in~\cite{eosk}, and discussed in \cite{pearl}, 
there are already changes as the universality scale
is lowered to $M_{in} = 10^{14}$ GeV.
The allowed focus-point region starts to separate from the LEP
chargino bound, moving to larger $m_{1/2}$. 
For $M_{in} = 10^{13}$ GeV,  we found
that  the allowed focus-point region also dips
further down, away from the electroweak vacuum condition boundary, while the
coannihilation strip moves up and farther away from the region where the
stau is the LSP.  Another remarkable feature at this value of
$M_{in}$ is the appearance of the rapid-annihilation funnel, familiar 
in the GUT-scale CMSSM at
large $\tan \beta$, but an unfamiliar feature for $\tb = 10$. 

As the universality scale is further decreased to $M_{in} = 10^{12.5}$~GeV,
as shown in panel (a) of Fig.~\ref{fig:mint}, 
the atoll formed by the conjunction of what had been the focus-point and
coannihilation strips has shrunk, so that it lies entirely within the range of
$(m_{1/2}, m_0)$ shown in panel (a).
We now see clearly two distinct
regions of the plane excluded due to an excess relic density of
neutralinos; the area enclosed by the atoll and the slice between the
lower funnel wall and the boundary of the already-excluded $\stau$-LSP
region.

 In panel (b) of of Fig.~\ref{fig:mint} for $M_{in} = 10^{12}$~GeV, 
the focus-point and coannihilation regions
are fully combined and the atoll has mostly filled in to become a small island of acceptable relic
density.  To the right of this island is a strip that is provided by the
lower funnel wall.  The strip curves slightly as
$m_{1/2}$ increases then takes a sharp plunge back down towards the
boundary of the region where the stau is the LSP, a feature
associated with the $\chi \chi \to h + A$ threshold. Reduction in
the universality scale from this point results in the lower funnel wall being pushed down into the
excluded $\stau$ LSP region and total evaporation of
the island as seen in  Fig.~\ref{fig:mint2} and only a
small residual turquoise region at large $m_{1/2}$ where
the relic density is within the WMAP limits.  At all other points in the visible part
of the $(m_{1/2},m_0)$ plane the relic density of neutralinos is too low
to provide fully the cold dark matter density preferred by WMAP {\it et al}. 
Of course, these SUSY models
would not be excluded if there is another source of cold dark matter in the
universe.

\begin{figure}[ht!]
\includegraphics[height=3in]{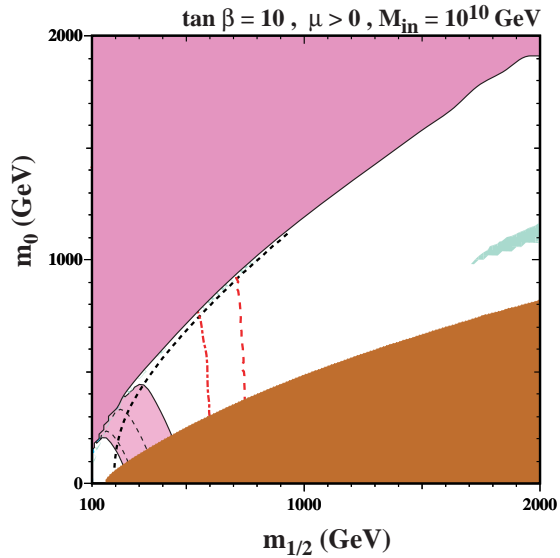}
\caption{\label{fig:mint2}
{\it 
As in Fig. \protect\ref{fig:mint} for $M_{in} = 10^{10}$ GeV. }}
\end{figure}

\section{mSUGRA models}

As discussed earlier, mSUGRA models, in contrast to the CMSSM,
have an additional boundary condition, namely the value of the 
bilinear $B$-term is fixed at the GUT scale to $B_0 = A_0 - m_0$.
As a consequence, one is no longer free to choose a
fixed value of $\tb$ across the $m_{1/2}-m_0$ plane.
Instead, $\tb$ is derived for each value of $m_{1/2}, m_0$ and $A_0$ \cite{vcmssm}.
Phenomenologically distinct planes are rather determined by
a choice for $A_0/m_0$.   In Fig. \ref{fig:msugra}, two such planes are shown
with (a) $A_0/m_0 = 3 - \sqrt{3}$ as predicted in the simplest model of
supersymmetry breaking \cite{pol} and with (b) $A_0/m_0 = 2.0$.
Shown are the contours of $\tan \beta$ (solid
blue lines) in the $(m_{1/2}, m_0)$ planes. 
Also shown are the contours where
$m_{\chi^\pm} > 104$~GeV (near-vertical black dashed lines) and $m_h >
114$~GeV (diagonal red dash-dotted lines). The regions excluded by $b \to
s \gamma$ have medium (green) shading, and those where the relic density
of neutralinos lies within the WMAP range have light (turquoise) shading. 
The region suggested by $g_\mu - 2$ at 2-$\sigma$ has very light (pink) shading.
As one can see, relatively low values of $\tb$ are obtained across the planes.

Another difference between the CMSSM and models based on
mSUGRA concerns the mass of the gravitino.
In the CMSSM, it is not specified and and can be taken suitably large so that the 
neutralino is the LSP (outside the stau LSP region). In mSUGRA, the scalar masses,
$m_0$ are determined by (and equal to) the gravitino mass. In Fig. \ref{fig:msugra}, 
the gravitino LSP and the neutralino LSP regions are separated by dark (chocolate)
solid lines.
Above this line, the neutralino (or stau) is the LSP, whilst below it, the gravitino is the LSP
\cite{gdm}.
As one can see by comparing the two panels, the potential for neutralino dark matter
in mSUGRA models is dependent on $A_0/m_0$.  In panel (a), the only areas 
where the neutralino density are not too large occur where the Higgs mass
is far too small or, at higher $m_0$ the chargino mass is too small. 
At larger $A_0/m_0$, the co-annihilation strip rises above the
neutralino-gravitino delineation.  In panel (b), we see the familar co-annihilation strip.
It should be noted that the focus point region is not realized in mSUGRA models
as the value of $\mu$ does not decrease with increasing $m_0$ when $A_0/m_0$ is 
fixed and $B_0 = A_0 - m_0$. There are also no funnel regions, as $\tb$ is never sufficiently high.

\begin{figure}[h]
\includegraphics[height=3in]{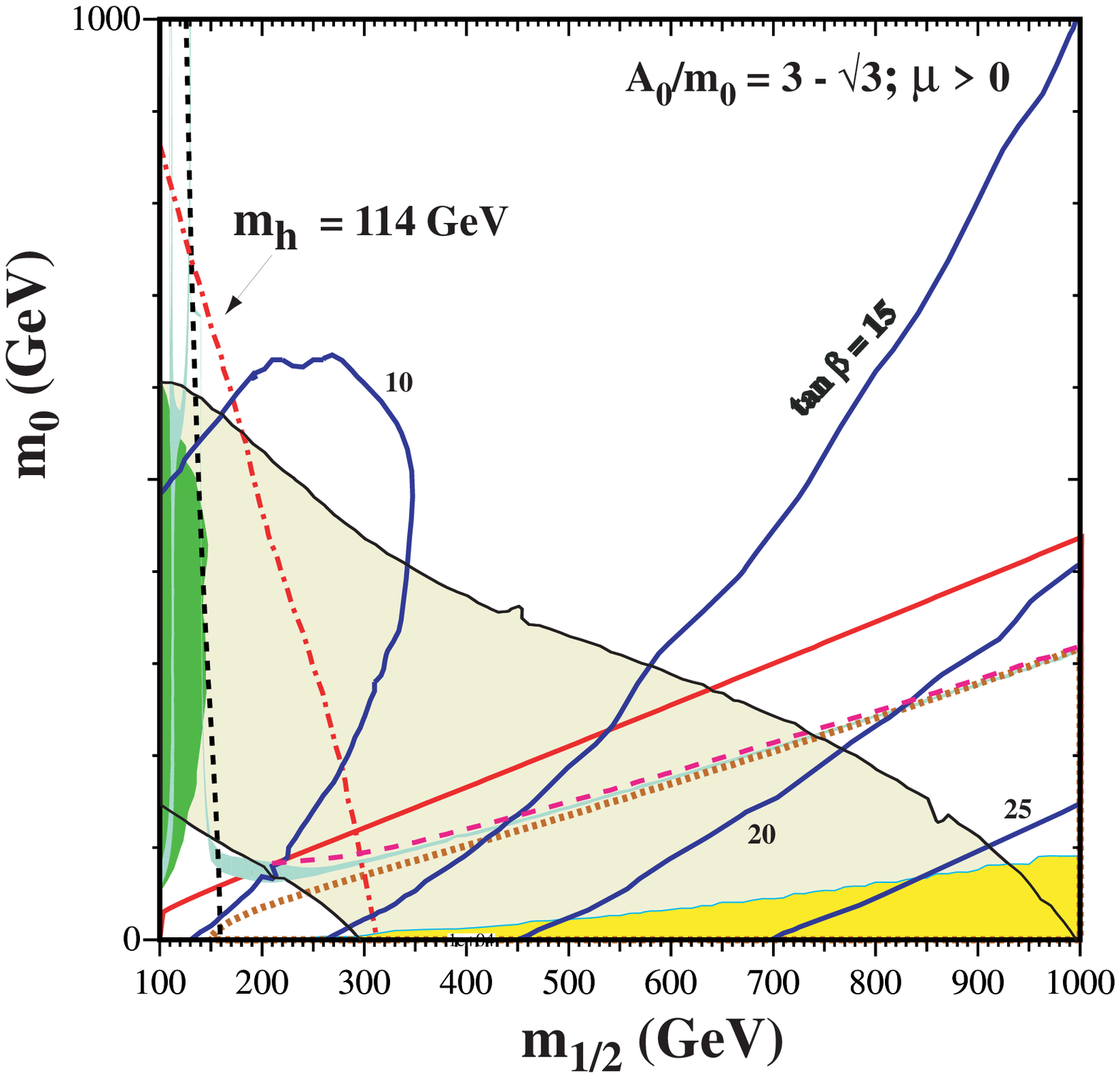}
\includegraphics[height=3in]{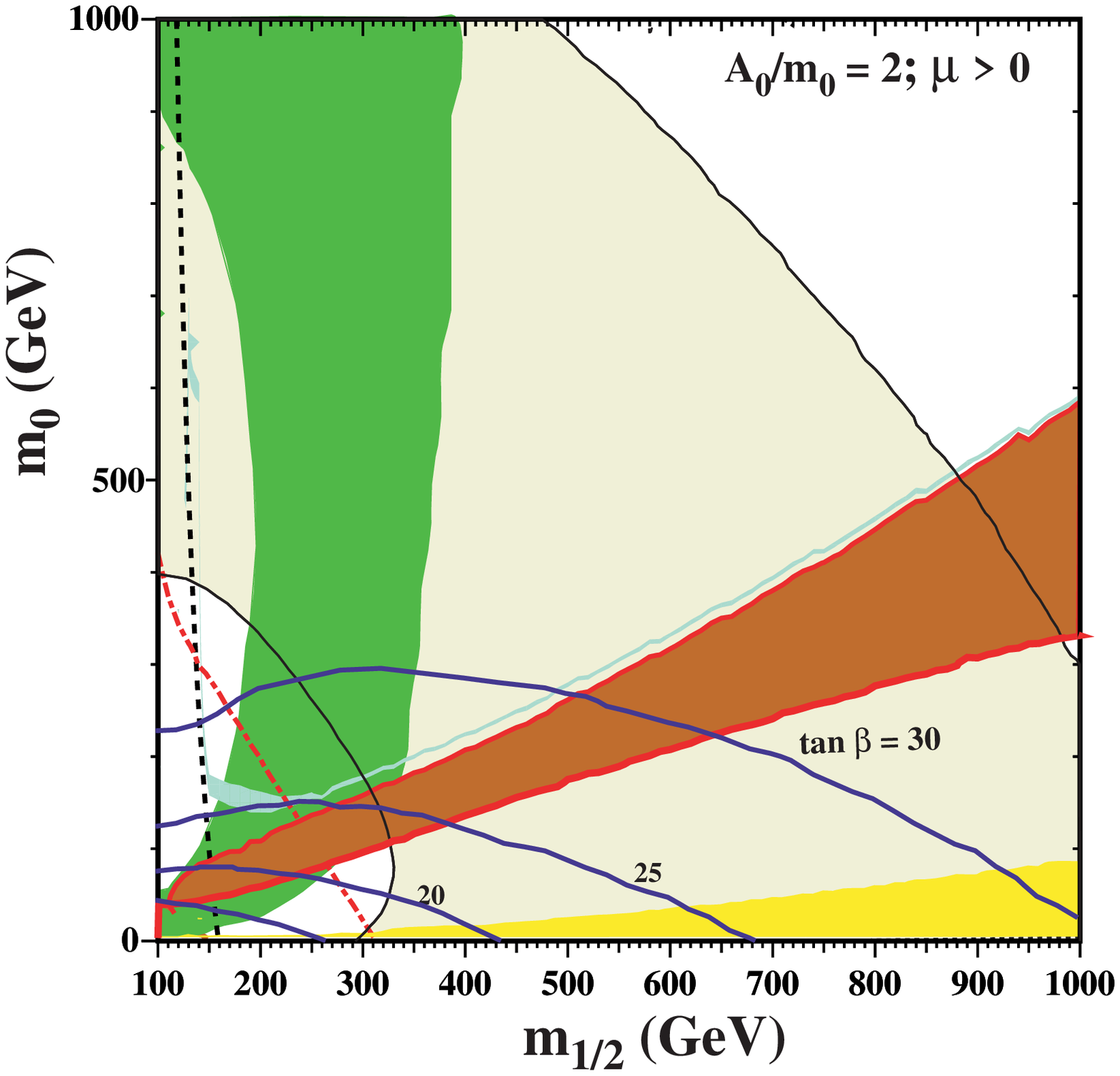}
\caption{\label{fig:msugra}
{\it 
Examples of mSUGRA $(m_{1/2}, m_0)$ planes with contours of $\tan \beta$ 
superposed, for $\mu > 0$ and (a) the simplest Polonyi model with $A_0/m_0 = 3 - 
\sqrt{3}$,  and (b) $A_0/m_0 = 2.0$, all with $ B_0 =
A_0 -m_0$. In each panel, we show the regions excluded by 
the LEP lower limits on MSSM particles and those ruled out by $b
\to s \gamma$ decay (medium green shading): the regions 
favoured by $g_\mu - 2$ are very light (yellow) shaded, bordered by a thin
(black) line.
The dark (chocolate) solid lines separate the 
gravitino LSP regions. The regions favoured 
by WMAP in the neutralino LSP case have light (turquoise)
shading. The dashed (pink) line corresponds to the maximum relic density 
for the gravitino LSP, and regions allowed by BBN constraint neglecting the effects of bound states on NSP 
decay are light (yellow) shaded.}}
\end{figure}

In the gravitino LSP regions, the next-to-lightest supersymmetric particle (NSP)
may be either the neutralino or stau which is now unstable. 
The relic density of gravitinos is acceptably low only below the
dashed (pink) line. This excludes a supplementary domain of the $(m_{1/2},
m_0)$ plane in panel (a). 
However, the strongest constraint is provided by the effect of neutralino or stau
decays on big bang neucleosynthesis (BBN) \cite{decay1}.  
Outside the light (yellow) shaded region,
the decays spoil the success of BBN.  Note that in panel (b), there remains a
region which is excluded because the stau is the LSP. 

Recently, new attention has been focussed on the regions with a stau
NSP due to its ability to form bound states (primarily with $^4$He). 
When such bound states occur, they catalyze certain nuclear
reactions such as $^4$He(D, $\gamma$)$^6$Li which is normally highly suppressed
due to the production of a low energy $\gamma$ whereas the bound state reaction is not
\cite{decay2}. In Fig. \ref{fig:cefos1}a,  
the $(m_{1/2}, m_0)$ plane is displayed
showing explicit element abundance contours \cite{cefos} 
when the gravitino mass is  $m_{3/2} = 0.2 m_0$ in the 
{\it absence}
of stau bound state effects.
To the left of the solid black line 
the gravitino is the not the LSP.
The diagonal red dotted line corresponds to the boundary between
a neutralino and stau NSP.  Above the line, the neutralino is the NSP,
and below it, the NSP is the stau.  Very close to this boundary,
there is a diagonal brown solid line.  Above this line, the relic
density of gravitinos from NSP decay is too high, i.e.,
\beq
\frac{m_{3/2}}{m_{NSP}} \Omega_{NSP} h^2 > 0.12.
\eeq
Thus we should restrict our attention to the area below this line.

The very thick green line labelled \li7 = 4.3
corresponds to the contour where \li7/H = $4.3 \times 10^{-10}$, a value very
close to the standard BBN result for \li7/H. 
It forms a `V' shape, whose right edge runs along the neutralino-stau NSP border.
Below the V, the abundance of \li7 is smaller than the standard BBN result. 
However, for relatively small values of $m_{1/2}$, 
the \li7 abundance does not differ very much from this standard BBN result:
it is only when $m_{1/2} \ga 3000$~GeV that \li7 begins to drop significantly.  
The stau lifetime drops with increasing $m_{1/2}$, and when 
$\tau \sim 1000$ s, at $m_{1/2} \sim 4000$ GeV, the \li7 abundance has been reduced 
to an observation-friendly value close to $2 \times 10^{-10}$ as claimed in~\cite{jed}
and shown by the (unlabeled) thin dashed (green) contours.

The region where the
\li6/\li7 ratio lies between 0.01 and 0.15 forms a band which moves
from lower left to upper right.  As one can see in the orange
shading, there is a large region where the lithium isotopic ratio can be
made acceptable. However, if we restrict to D/H $< 4.0 \times 10^{-5}$, we
see that this ratio is interesting only when \li7 is at or slightly below
the standard BBN result.

Turning now to Fig.~\ref{fig:cefos1}b, we show the analogous results when
the bound-state effects are included in the calculation.  The abundance
contours are identical to those in panel (a) above the
diagonal dotted line, where the NSP is a neutralino and bound states do
not form.  We also note that the bound state effects on D and \he3 are
quite minimal, so that these element abundances are very similar to those
in Fig.~\ref{fig:cefos1}a.  However, comparing panels (a) and (b), one sees
dramatic bound-state effects on the lithium abundances.  
Everywhere to the left of the solid blue line labeled 0.15 is excluded.  
In the stau NSP region, this means that $m_{1/2} \ga 1500$~GeV.  
Moreover, in the stau region to the right of the \li6/\li7 = 0.15 contour,
the \li7 abundance drops below $9 \times 10^{-11}$ (as shown by the thin
green dotted curve).
In this case, not only do the bound-state effects
increase the \li6 abundance when $m_{1/2}$ is small (i.e., at relatively
long stau lifetimes), but they also decrease the \li7 abundance when the
lifetime of the stau is about 1500~s. Thus, at $(m_{1/2}, m_0) \simeq
(3200,400)$, we find that \li6/\li7 $\simeq 0.04$, \li7/H $\simeq 1.2
\times 10^{-10}$, and D/H $\simeq 3.8 \times 10^{5}$.  Indeed, when
$m_{1/2}$ is between 3000-4000 GeV, the bound state effects cut the \li7
abundance roughly in half. In the darker (pink) region, 
the lithium abundances match the observational plateau
values, with the properties $\li6/\li7 > 0.01$ and $0.9 \times10^{-10} <
\li7/\textrm{H} < 2.0 \times10^{-10}$.

 \begin{figure}[h]
\includegraphics[height=3in]{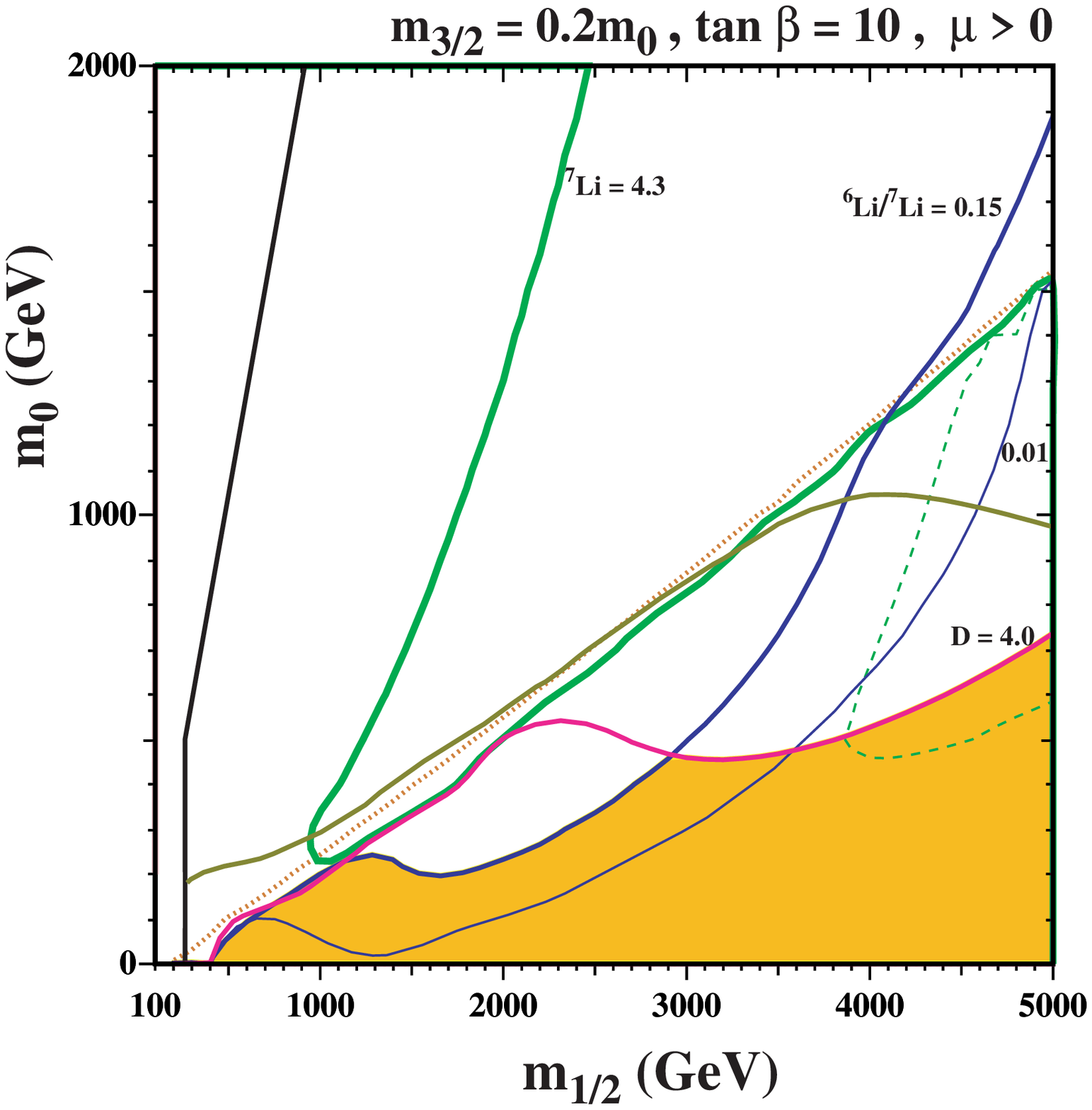}
\includegraphics[height=3in]{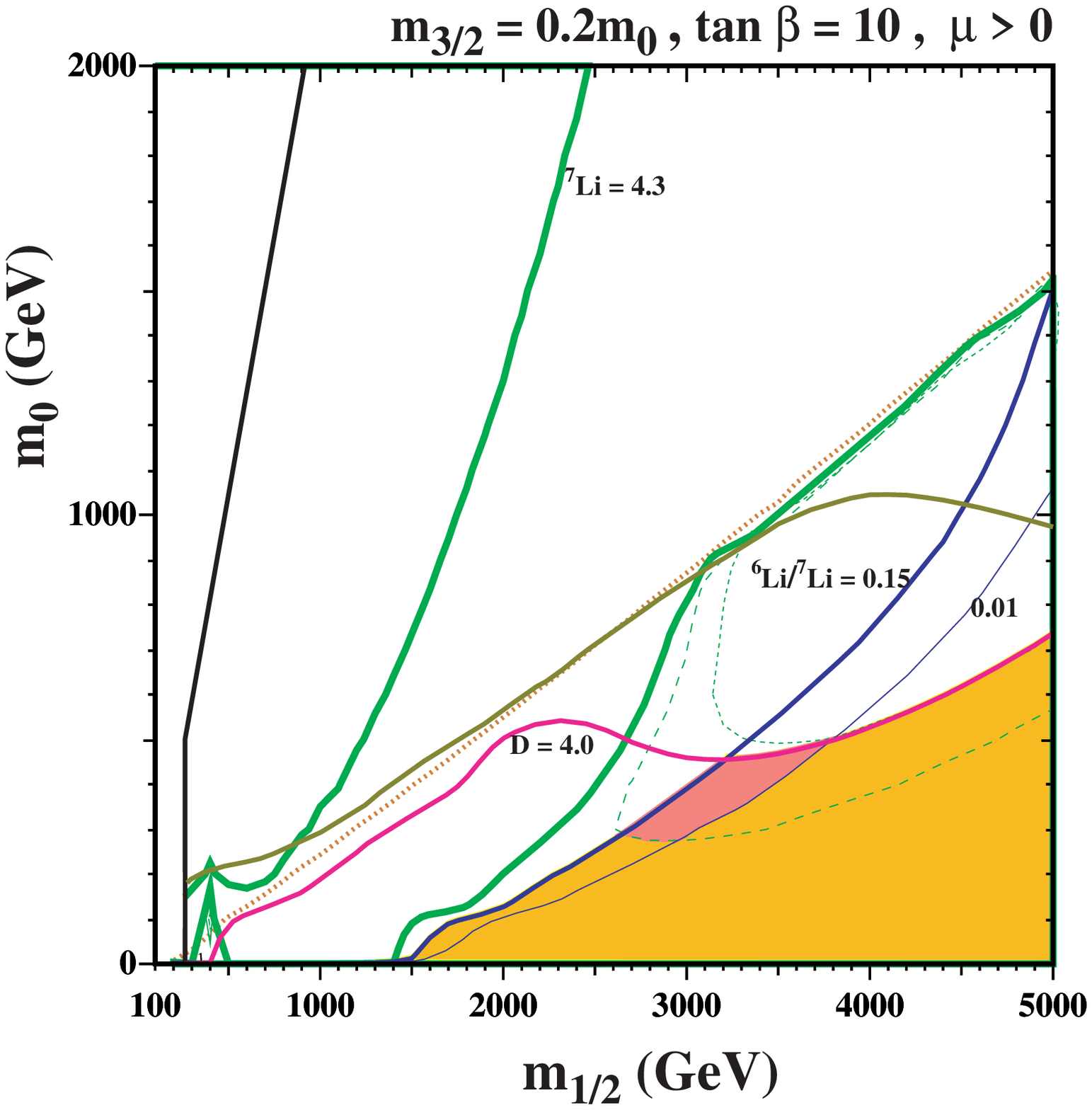}
\caption{\label{fig:cefos1}
{\it 
An$(m_{1/2}, m_0)$ planes for $A_0=0$, $\mu > 0$ and $\tan
\beta = 10$ with $m_{3/2} = 0.2 m_0$ (a) without and (b) with the effects of
boundstates included.
The regions to the left of the solid black lines are not considered, 
since there the gravitino is not the LSP.
In the orange (light) shaded regions, the differences between the calculated 
and observed light-element abundances are no greater than in standard 
BBN without late particle decays. In the pink (dark) shaded region in 
panel b, the 
abundances lie within the ranges favoured by observation. 
The significances of the 
other  lines and contours  are explained in the text.}}
\end{figure}

Next we return to an example of an mSUGRA model based on the Polonyi model for which
$A_0/m_0 = 3- \sqrt{3}$ with the condition that $m_{3/2}
= m_0$.  In Fig.~\ref{fig:cefos2}a, we show the mSUGRA model without the
bound states. In the upper part of the plane, we do not have gravitino dark matter.  We see
that \he3/D \cite{sigl} eliminates all but a triangular area which extends up to $m_0
= 1000$ GeV, when $m_{1/2} = 5000$ GeV.  Below the \he3/D = 1 contour, D
and \li7 are close to their standard BBN values, and there is a
substantial orange shaded region. We note that \li6 is interestingly high,
between 0.01 and 0.15 in much of this region.

As seen in Fig.~\ref{fig:cefos2}b, when bound-state effects are included
in this mSUGRA model, both lithium isotope abundances are too large except
in the extreme lower right corner, where there is a small region shaded 
orange. However, there is no region where the lithium abundances fall 
within the favoured plateau ranges.

\begin{figure}[h]
\includegraphics[height=3in]{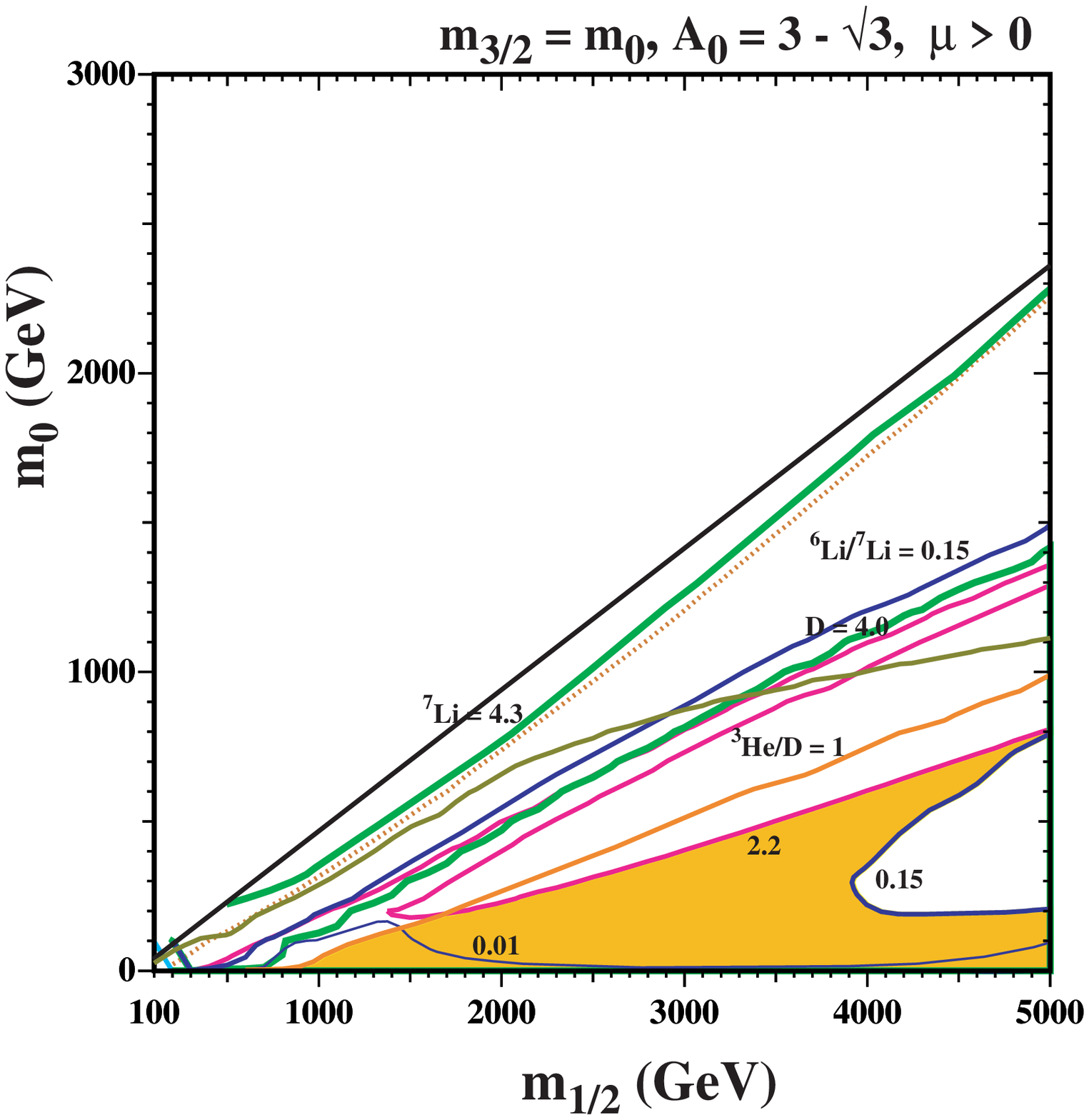}
\includegraphics[height=3in]{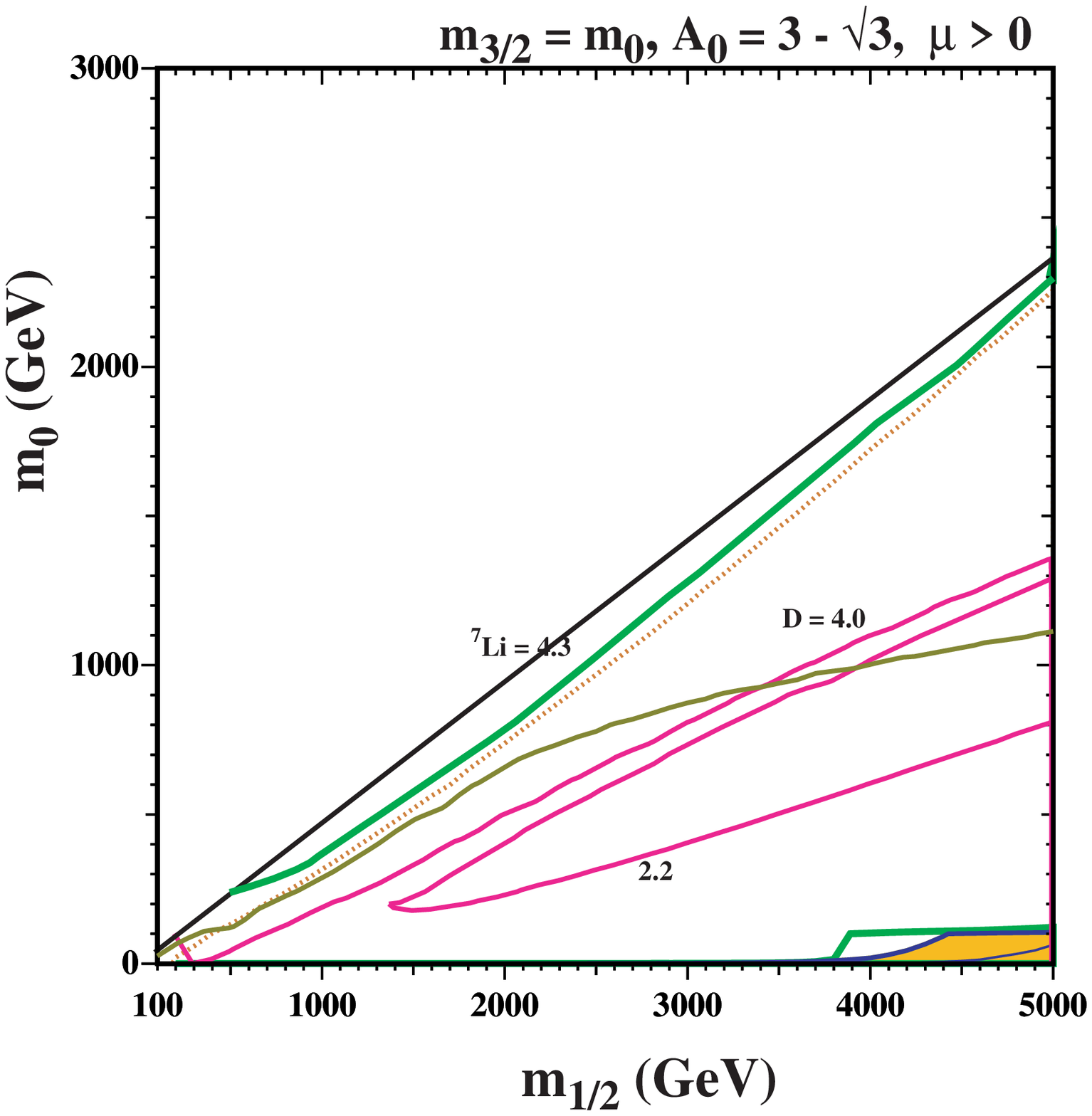}
\caption{\label{fig:cefos2}
{\it 
As in Fig. \protect{\ref{fig:cefos1}}, for the mSUGRA case with $A_0/m_0 = 3 - \sqrt{3}$.
}}
\end{figure}

The prospects for direct detection in the mSUGRA models (when the neutralino is the
LSP)  is very similar to that of the CMSSM but slightly more difficult. In a plot such as that shown
in Fig. \ref{fig:Andyall}, the mSUGRA predictions are down by a factor of about 2 or less.

\section{The NUHM}

In the NUHM, the Higgs soft masses are treated independently from $m_0$.
In effect, this allows one to choose $\mu$ and the Higgs pseudoscalar mass, $m_A$
freely (up to phenomenological constraints). Two examples of planes in the NUHM 
are shown in Fig. \ref{fig:nuhm}. In panel (a), an $m_{1/2}-m_0$ plane is shown
with $\mu = 700$~GeV and $m_A
= 400$~GeV fixed across the plane \cite{nuhm}. 
As usual, the light (turquoise) shaded area is
the cosmologically preferred region.
There is a bulk region satisfying this preference at $m_{1/2} \sim 150$~GeV
to 350~GeV and $m_0 \sim 100$~GeV. 
The dark (red) shaded regions are excluded because a charged
sparticle is lighter than the neutralino.  As in the CMSSM, 
there are light (turquoise) shaded strips close to
these forbidden regions where coannihilation suppresses the relic density
sufficiently to be cosmologically acceptable. Further away from these
regions, the relic density is generally too high. 
At small $m_{1/2}$ and $m_0$ the left handed
sleptons, and also the sneutrinos, become lighter than the neutralino. The
darker (dark blue) shaded area is where a sneutrino is the LSP. 

The near-vertical dark (black) dashed and light (red) dot-dashed lines in
Fig.~\ref{fig:nuhm} are the LEP exclusion contours $m_{\chi^\pm} > 104$~GeV
and $m_h > 114$~GeV respectively. As in the CMSSM case, they exclude low values of
$m_{1/2}$, and hence rule out rapid relic annihilation via direct-channel
$h$ and $Z^0$ poles. The solid lines curved around small values of
$m_{1/2}$ and $m_0$ bound the light (pink) shaded region favoured by
$a_\mu$ and recent analyses of the $e^+ e^-$ data.

A striking feature in Fig.~\ref{fig:nuhm}(a) when $m_{1/2} \sim 450$ GeV is a
strip with low $\ohsq$, which has bands with acceptable relic density on
either side.  The low-$\ohsq$ strip is due to rapid annihilation via the
direct-channel $A, H$ poles which occur when $m_\chi = m_A / 2 = 200$~GeV,
indicated by the near-vertical solid (blue) line. These were found in the CMSSM
\cite{funnel,efgosi}, but at larger $\tan \beta$. 
At higher $m_{1/2} \sim 1300 - 1400$~GeV, there is another region of acceptable 
relic density. This band, the transition band,  is broadened because the neutralino acquires
significant Higgsino content as $m_{1/2}$ becomes greater than $\mu$, and the relic density is suppressed by the
increased $W^+ W^-$ production. To the right of this band, the relic density falls below the
WMAP value.

\begin{figure}[ht!]
\includegraphics[height=3in]{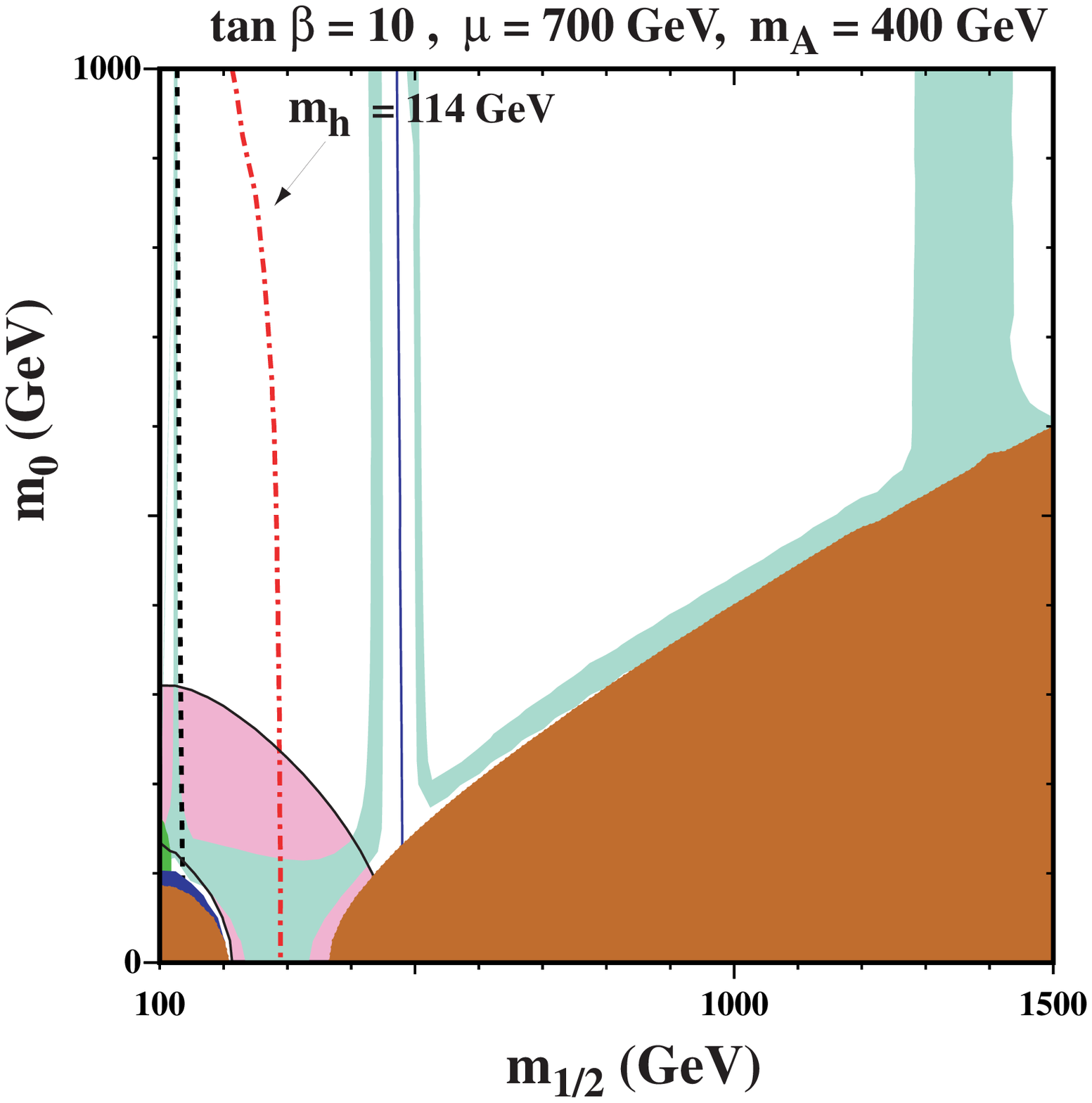}
\includegraphics[height=3in]{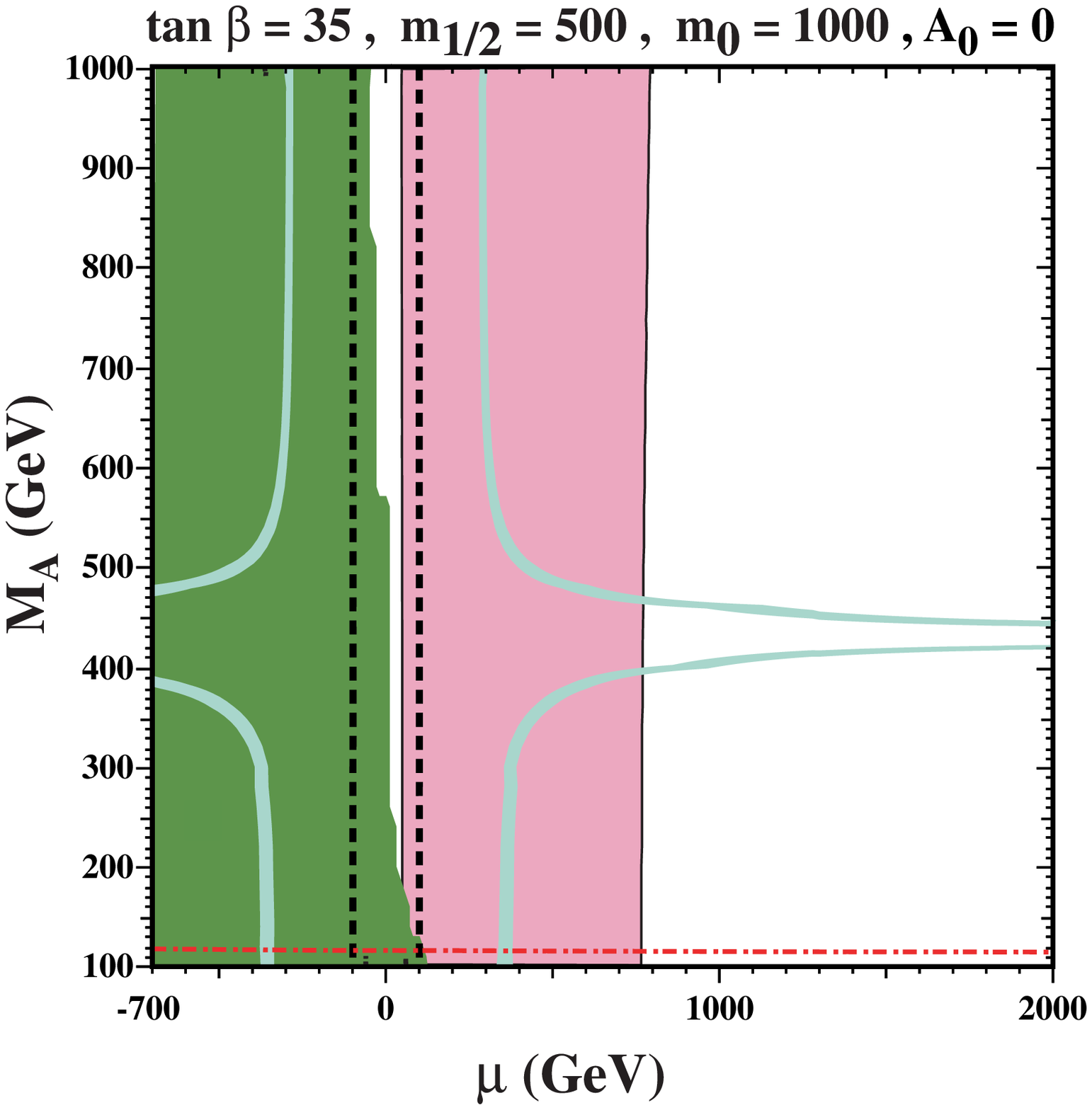}
\caption{\label{fig:nuhm}
{\it 
Projections of the NUHM model on (a) the $(m_{1/2}, m_0)$ planes for $\tan
\beta =  10$ and $\mu = 700$~GeV and $m_A
= 400$~GeV and (b) the $\mu-m_A$ plane with $m_{1/2} = 500$ GeV, 
$m_0 = 1000$ GeV,
$\tb = 35$
and $A_0 = 0$.  The (red)
dot-dashed lines are the contours $m_h = 114$~GeV as calculated using
{\tt FeynHiggs}~\cite{FeynHiggs}, and the near-vertical (black) dashed
lines are the contours $m_{\chi^\pm} = 104$~GeV. The
light (turquoise) shaded areas are the cosmologically preferred regions. 
The dark (brick red) shaded
regions is excluded because a charged particle is lighter than the 
neutralino,
and the darker (dark blue) shaded regions is excluded because the LSP is a
sneutrino. In panel (a) there is a very small medium
(green) shaded region excluded by $b \to s \gamma$, at small $m_{1/2}$. In panel (b),
it excluded most of $\mu < 0$.
The regions allowed by the E821
measurement of $a_\mu$ at the 2-$\sigma$ level, 
are shaded (pink) and bounded by solid black lines.
}}
\end{figure}

Another example of an NUHM plane is shown in panel (b) of Fig. \ref{fig:nuhm},
representing an $\mu-m_A$ plane with
a fixed choice of values of $\tan \beta = 35$,
$m_{1/2} = 500$ GeV, $m_0 = 1000$ GeV and $A_0 = 0$ \cite{eoss8}.
Line types and shading are as in panel (a). The region
with acceptable relic density now takes the form of a 
narrow strip of values of $\mu \sim  \pm 300 \to 350 \gev$ where 
the relic density lies within the WMAP range for almost all values of $m_A$.
The exception is a narrow strip centred on $m_A \sim 430 \gev$,
namely the rapid-annihilation funnel where
$m_\chi \sim m_A/2$, which would be acceptable if there is some other
source of cold dark matter. 
Note the increased importance of the $b \to s \gamma$ constraint which 
now excludes almost all of $\mu < 0$. 

By using the properties found in the $\mu-m_A$ plane discussed above,
one can construct a $m_A-\tan \beta$ plane such that (nearly) the entire
plane respects the cosmological relic density constraint \cite{ehoww,ehhow}.
These plane can be used as a particularly useful benchmark for
existing and future searchs for Higgs bosons \cite{ehhow}.
An example of such a plane is shown in Fig. \ref{fig:ehhow}
with fixed $m_{1/2} = 500$ GeV and $m_0 = 1000$ GeV. As one can see, the
light shaded (turquoise) region now fills up the plane except for a 
vertical swath where s-channel annihilation through $m_A$ drives
the relic density to small values. 
A likelihood analysis of this NUHM benchmark surface, including the
EWPO $\MW$, $\sweff$, $\Ga_Z$, $(g-2)_\mu$
and $\Mh$ and the BPO 
$\br(b \to s \gamma)$, $\br(B_s \to \mu^+\mu^-)$, 
$\br(B_u \to \tau \nu_\tau)$ and $\De M_{B_s}$ was performed recently
in \cite{ehoww}. The lowest $\chi^2$ value in this plane is $\chi^2_{\rm min} = 7.4$ and
the corresponding best-fit point is shown by a (red) cross as are 
the $\Delta \chi^2 = 2.30$ and 4.61 contours around the best-fit point.
The (black shaded) region in the lower left is
excluded  at the 95~\% C.L.\ by the LEP Higgs
searches in the channel
$e^+e^- \to Z^* \to Z h, H$~\cite{LEPHiggs}.

\begin{figure}[ht!]
\includegraphics[height=3in]{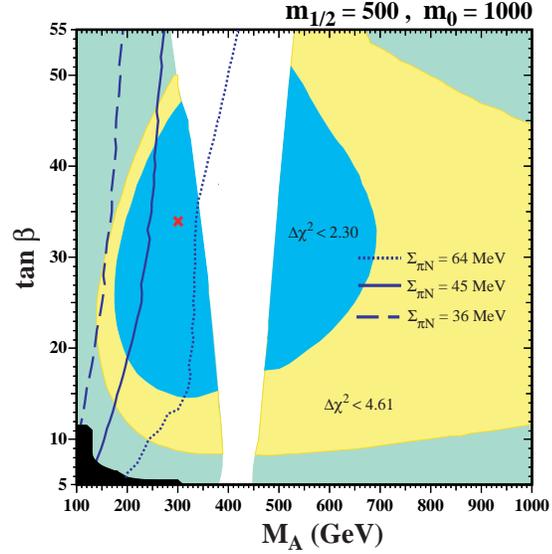}
\caption{\label{fig:ehhow}
{\it The $(m_A, \tb)$ plane for the NUHM benchmark surface 
displaying the contours of $\Delta \chi^2$
found in a recent global fit to EWPO and BPO~\protect\cite{ehoww}.
Also displayed is the expected sensitivity of
present direct searches for the scattering
of dark matter particles. }}
\end{figure}

Also shown in Fig. \ref{fig:nuhm} is the current reach of the 
XENON10 experiment~\cite{xenon}.
The solid line in the figure correspond to the XENON10 bound obtained
assuming $\Sigma_{\pi N} = 45$ MeV.   The corresponding reach obtained if
$\Sigma_{\pi N} = 64$ MeV is shown by the dotted curve. As one can see, in this 
case, the best fit point would already have been probed by the XENON10 experiment.
If the strangeness content of the proton were zero, it would correspond to $\Sigma_{\pi N} = 36$ MeV,
and the XENON10 reach in this case is displayed by the dashed curve.
Finally, we note that a future  
sensitivity to a cross section of $10^{-8}$ pb would cover the
entire surface.

As can be inferred from the benchmark surface shown above, in the NUHM,
current constraints already exclude many interesting models \cite{eoss8}.
These models generically
have large scattering cross sections.  
Scatter plots of showing the reach of CDMS and XENON10 for both 
$\Sigma_{\pi N} = 45$ MeV and 64 MeV are shown in Fig. \ref{fig:nuhmdd}.
In comparison with Fig. \ref{fig:Andyall}, one sees that many more viable models
are already currently being probed.

\begin{figure}[ht!]
\centering
\includegraphics[height=3in]{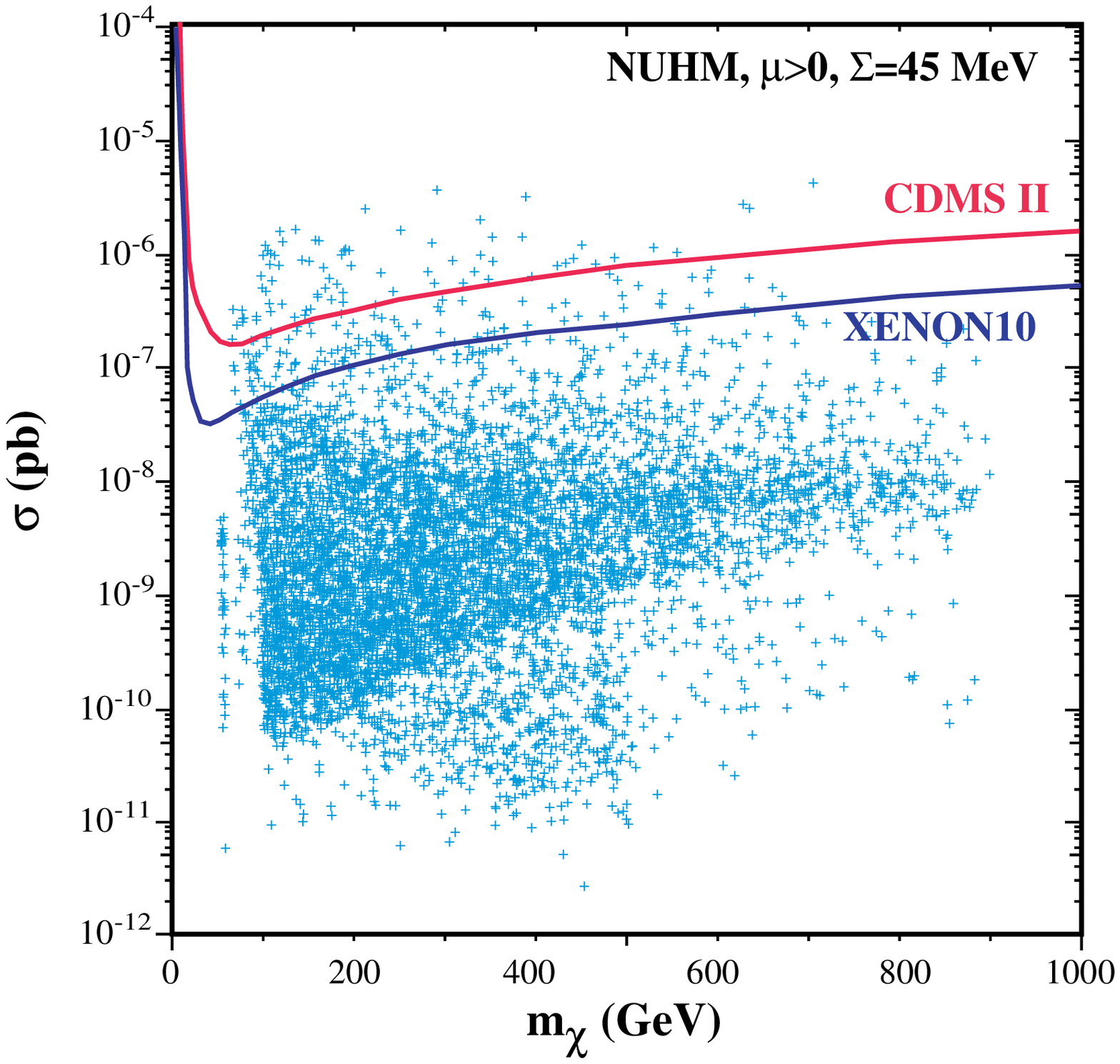}
\includegraphics[height=3in]{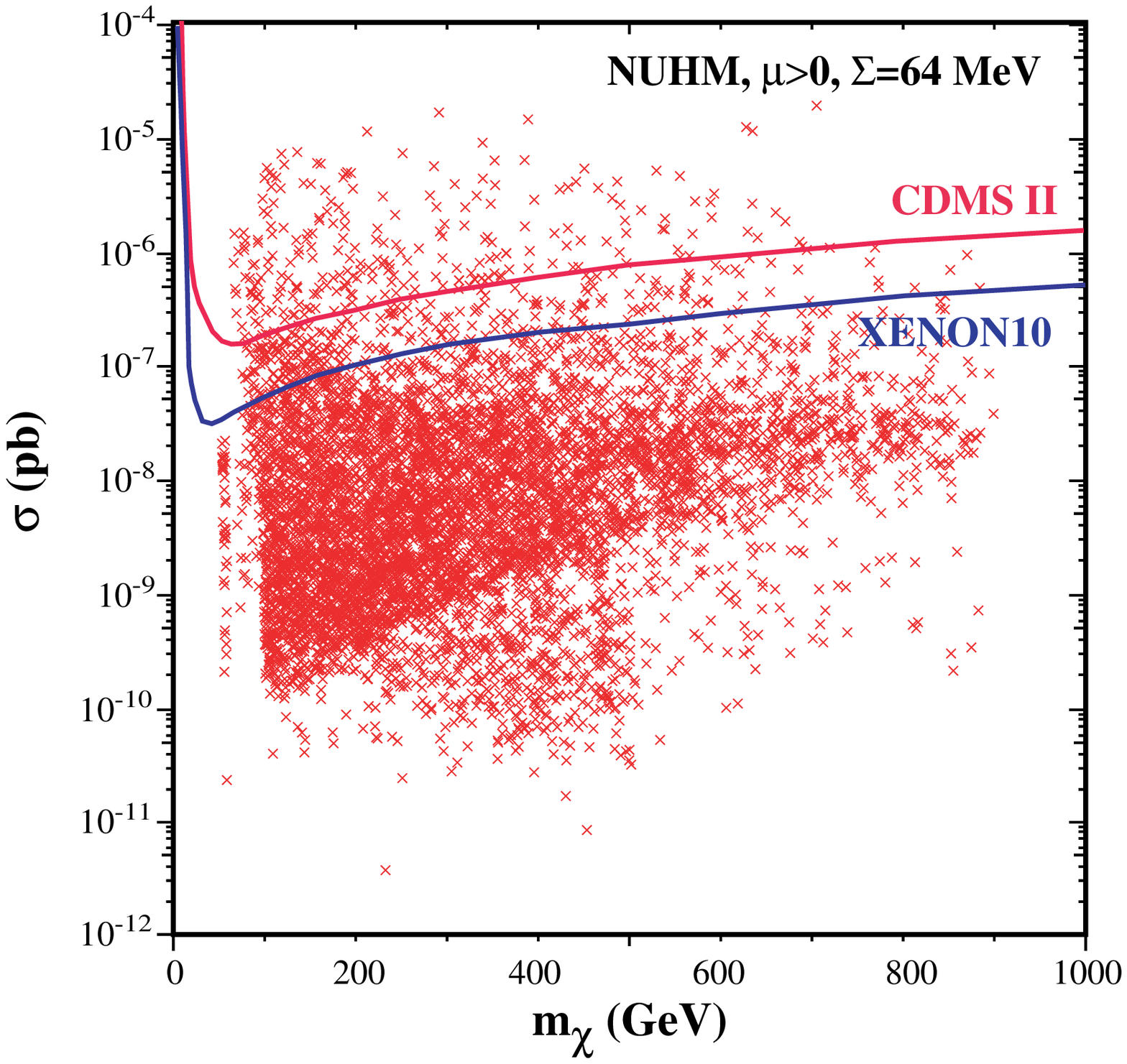}
\caption{\label{fig:nuhmdd}
{\it 
Scatter plots of the spin-independent elastic-scattering cross
section predicted in the CMSSM (a) for $\Sigma_{\pi N} = 45$ MeV (b) and 64 MeV.}}
\end{figure}

\section{A Hint of Higgs}

The CDF and D0 Collaborations have already
established important limits on the (heavier) MSSM Higgs bosons,
particularly at large
$\tb$~\cite{CDFbounds,D0bounds,Tevcharged,CDFnew,D0new}.
Recently the CDF Collaboration, investigating the channel
\beq
p \bar p \to \phi \to \tau^+\tau^-, \quad (\phi = h,H,A)~,
\label{pphtautau}
\eeq
has been unable to improve these
limits to the extent of the sensitivity expected with the analyzed integrated
luminosity of $\sim 1 \ifb$~\cite{CDFnew}, whereas there is no indication of 
any similar effect in D0 data~\cite{D0new}.%
Within the MSSM the channel (\ref{pphtautau}) is enhanced as compared
to the corresponding SM process by roughly a factor of 
$\TQb/((1 + \Delta_b)^2 + 9)$~\cite{benchmark3}, where $\Delta_b$ includes loop corrections to
the $\phi b \bar b$ vertex (see \cite{benchmark3} for
details) and is subdominant for the $\tau^+\tau^-$ final state.
Correspondingly, the unexpected weakness of the CDF 
exclusion might be explicable within the MSSM if $m_A \approx 160 \gev$ 
and $\tb \ga 45$.

In the CMSSM, there are no parameter sets consistent
with all the experimental and theoretical constraints.
In particular, low values of $m_{1/2}$ are required to obtain low values of $m_A$.  
For $m_A \simeq 160 \gev$ and the other parameters
tuned to fulfill the $B$~physics constraints, there are no CMSSM 
solutions for which the lightest Higgs mass is not in direct contradiction
with experimental limits~\cite{LEPHiggs}.
However, we do find that all the constraints may be
satisfied in the NUHM \cite{NUHMad,ehow5}.  

We consider first the \plane{m_{1/2}}{m_0} shown in panel (a) of Fig.
\ref{fig:hint1}, which has $\tb = 45$, $\mu = 370 \gev$ and
$A_0 = -1800 \gev$, as well as $\MA = 160 \gev$. 
As before, the dark (brown) shaded region at low $m_0$ is forbidden, because
there the LSP would be the lighter
stau. The WMAP cold dark matter constraint is satisfied only within the
lighter (turquoise) shaded region. 
To the left of this region, the relic density is too small, due to $s$-channel
annihilation through the Higgs pseudoscalar $A$. As $m_{1/2}$ increases away
from the pole, the relic density increases toward the WMAP range.
However, as $m_{1/2}$ is increased, the neutralino 
acquires a larger Higgsino component and annihilations to pairs of
$W$~and $Z$~bosons become enhanced. To the right of this transition
region, the relic density again 
lies below the WMAP preferred value.  The shaded region here is therefore
an overlap of the funnel and transition regions discussed in~\cite{nuhm}.
The $\br(B_s \to \mu^+\mu^-)$ constraint \cite{bmmex} is satisfied between the outer 
black dash-dotted lines, labelled $10^{-7}$, representing the
current limit on that branching ratio. 
Also shown are the contours where the
branching ratio is $2 \times 10^{-8}$, close to the sensitivity likely to be
attainable soon by CDF and D0.
Between these two contours, there is a strong cancellation
between  the flavor-violating contributions arising from the Higgs
and chargino couplings at the one-loop level and the Wilson coefficient
counterterms contributing to $\br(B_s \to \mu^+\mu^-)$ \cite{bmm}.

The dash-dotted (red) line shows the contour corresponding to
$\Mh = 114 \gev$, and only the region to the right of this line is
compatible with the constraint imposed by $\Mh$ (it should be kept in
mind that there is still a $\sim 3 \gev$ uncertainty in the
prediction of $\Mh$~\cite{FeynHiggs}).
Also shown in pink shading is the
region favoured by $(g-2)_\mu$ at the two-$\sigma$ 
level. The one- and two-$\sigma$ contours for $(g-2)_\mu$
are shown as elliptical dashed and solid black contours, respectively.
The region which is compatible with the WMAP relic density and $\Mh$,
and is also within the two-$\sigma$ $(g-2)_\mu$ experimental bound,
has $\br(B_s \to \mu^+\mu^-) > 2 \times 10^{-8}$. 
The measured value of $\br(b \to s \gamma)$ is in agreement with the theory
prediction only to the left
of the solid (green) region.

\begin{figure}[ht!]
\centering
\includegraphics[height=3in]{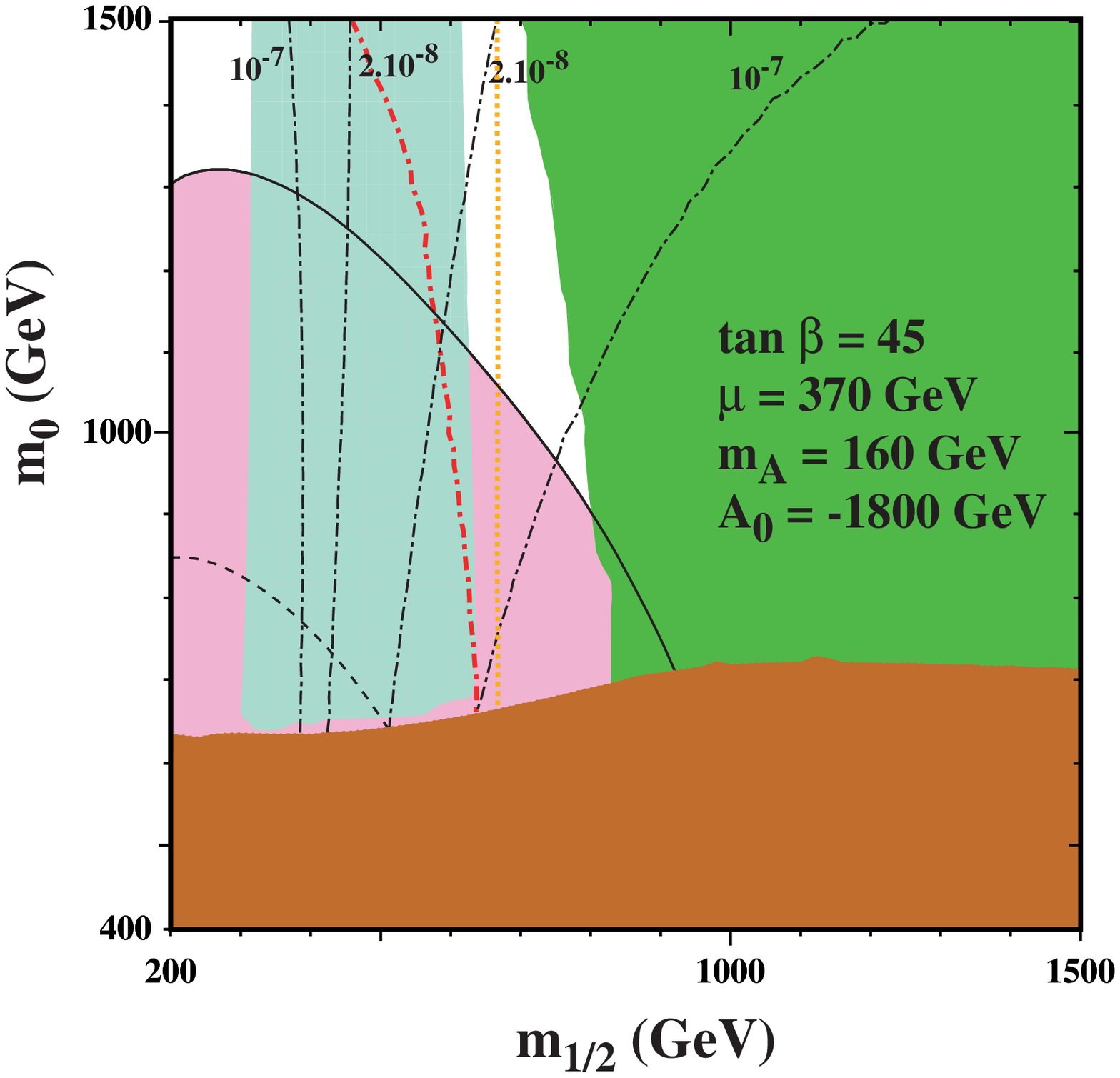}
\includegraphics[height=3in]{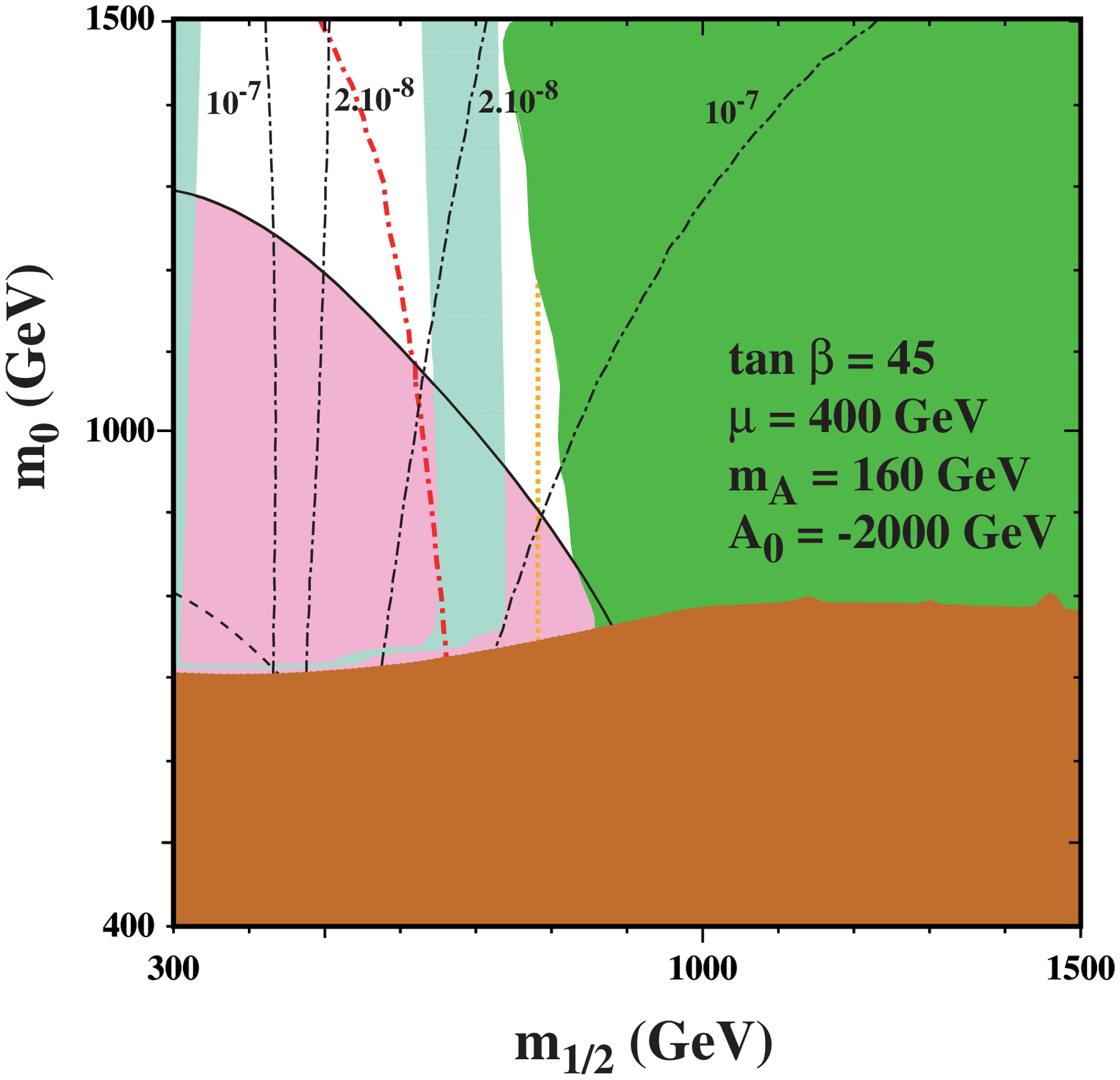}
\caption{\label{fig:hint1}
{\it 
The NUHM parameter space as a function of $m_{1/2}$ and $m_0$ for 
$\mu = 370 (400) \gev$ and $A_0 = -1800 (-2000) \gev$ in panels a (b). 
We fix $\MA = 160 \gev$, $\tb = 45$, $m_t = 171.4 \gev$
and $m_b = 4.25 \gev$. 
For the description of the various lines and shaded areas, see the text. }}
\end{figure}

We see that there is a narrow wedge of allowed parameter space in
\ref{fig:hint1}(a), which has $m_{1/2} \sim 600 \gev$ and 
$m_0 \sim 700$ to 1100~GeV. The $\br(b \to s \gamma)$ constraint is satisfied
easily throughout this region, and $(g-2)_\mu$ 
cuts off the top of the wedge, which would otherwise have extended to
$m_0 \gg 1500 \gev$. Within the allowed wedge, $\Mh$ is very close to
the LEP lower limit, and $\br(B_s \to \mu^+\mu^-) > 2 \times 10^{-8}$.
If  $\MA$ were much
smaller ($< 130 \gev$), there would be no wedge consistent 
simultaneously with the $\Omega_{\rm CDM}$,
$\Mh$ and $\br(B_s \to \mu^+\mu^-)$ constraints.

Given the matrix element uncertainties for direct detection summarized above, we show 
in \ref{fig:hint1}a) the 1-$\sigma$ lower limit on the calculated value of
the elastic cross section as compared to the CDMS upper limit~\cite{cdms}.
In the portion of the plane to the left
of the (orange) dotted line, the lower limit on the calculated 
spin-independent elastic cross section is smaller than the CDMS upper bound,
assuming the canonical local density. Whilst we have assumed
$\Sigma_{\pi N} = 45 $ MeV, the calculated lower limit effectively
assumes zero strangeness contribution to the proton mass, i.e., $y =
0$. In the region of interest, the lower limit 
on the calculated cross section is about 80\% of the CDMS upper bound, whereas
with a strangeness contribution of $y = 0.2$, the cross section 
would exceed the CDMS bound by a factor of $\sim 3$. 
Thus, if Nature has picked this corner of the NUHM parameter space, we expect 
direct detection of dark matter to be imminent.
Consistency with the XENON10 limit would further require a
reduction in the local dark matter density (to its lower limit). 
Intriguingly, the
XENON10 experiment has seen some potential signal events that are, however,
interpreted as background. 

Panel (a) of \ref{fig:hint2} explores the \plane{\mu}{A_0} for the choice
$(m_{1/2}, m_0)$ $= (600, 800) \gev$, values close to the lower tip of
the allowed wedge in \ref{fig:hint1}(a). In this case, the region
allowed by the $\br(B_s \to \mu^+\mu^-)$ constraint is {\it below} the
upper dash-dotted black 
line, and the LEP $\Mh$ constraint is satisfied only {\it above} the
dash-dotted red line. We see that only a restricted range 
$360 \gev < \mu < 390 \gev$ is compatible with the dark matter constraint.
This corresponds to the transition strip where the neutralino is the
appropriate bino/Higgsino combination.  To the left of this strip,
the relic density is too small and to the right, it is too large.
Only a very restricted range of $A_0 \sim - 1600 \gev$ is compatible
simultaneously with the $\Mh$ and $\br(B_s \to \mu^+\mu^-)$ constraints. 
Very large negative values of $A_0$ are excluded as the LSP is the lighter
stau.
On the \plane{\mu}{A_0}, the elastic scattering cross section
is a rapidly decreasing function of $\mu$ and is almost independent of $A_0$.
Indeed, NUHM points excluded by CDMS (or XENON10) generally have low values of 
$\mu$ and $\MA$~\cite{eoss}.
Values of $\mu > 355 \gev$ are compatible with CDMS if the strangeness
contribution to the proton mass is negligible.
 For this choice of parameters,
the entire displayed plane is compatible with $\br(b \to s \gamma)$ and
$(g-2)_\mu$.

\begin{figure}[ht!]
\centering
\includegraphics[height=3in]{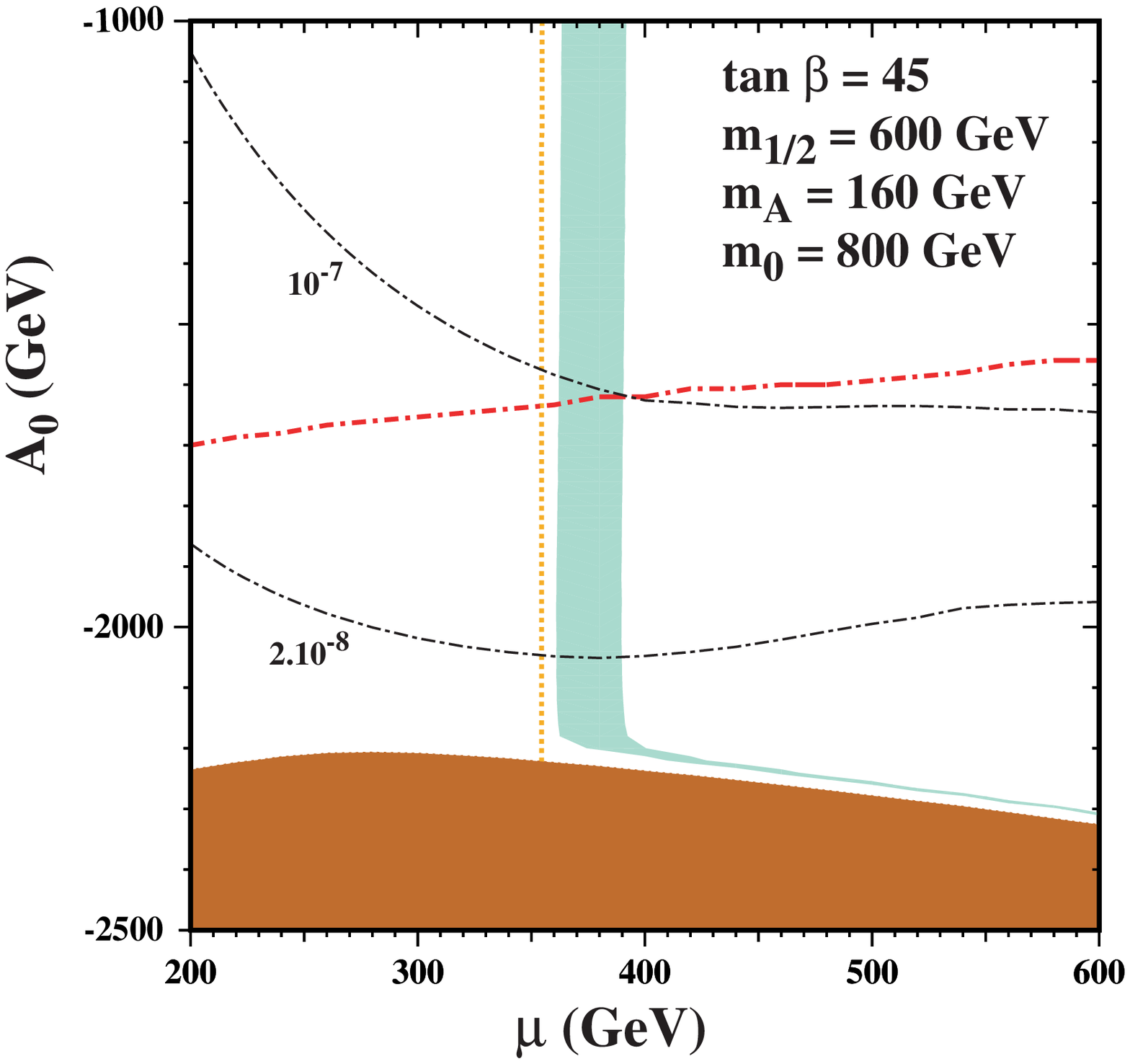}
\includegraphics[height=3in]{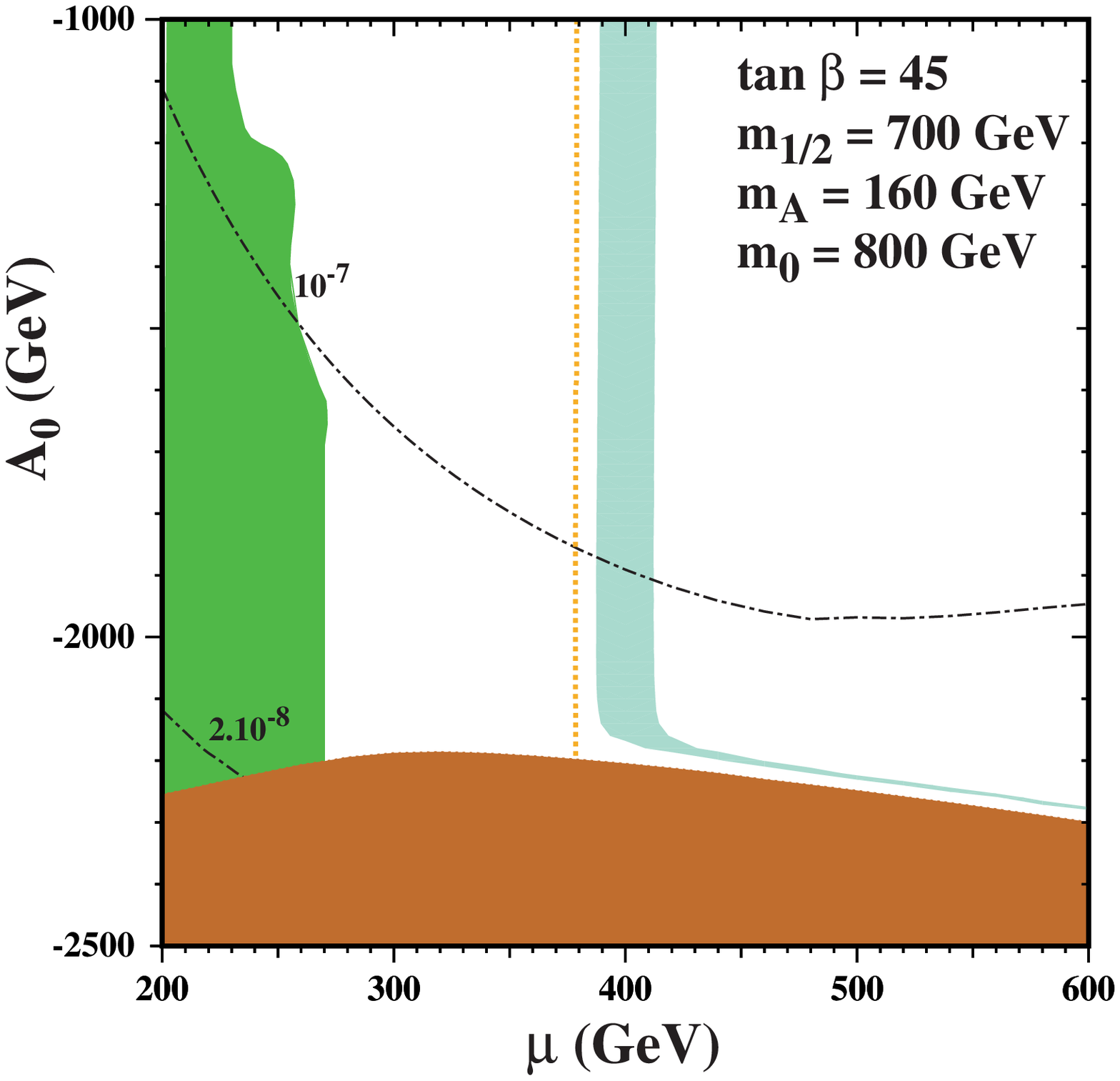}
\caption{\label{fig:hint2}
{\it 
The NUHM parameter space as a function of $\mu$ and $A_0$ for 
$m_{1/2} = 600 (700) \gev$ and $m_0 = 800 \gev$ in panels a (b).
We fix $\MA = 160 \gev$, $\tb = 45$, $m_t = 171.4 \gev$
and $m_b = 4.25 \gev$. 
For the description of the various lines and shaded areas, see the text.
 }}
\end{figure}

Panel (b) of Fig.
\ref{fig:hint2} shows what happens if $m_{1/2}$ is increased to
700~GeV, keeping $m_0$ and the other inputs the same.
Compared to Fig.~\ref{fig:hint2}(a), we see that the
WMAP strip becomes narrower and shifts to larger $\mu \sim 400 \gev$,  
and that $\br(b \to s \gamma)$ starts to
exclude a region visible at smaller $\mu$. If $m_{1/2}$ were to be
increased much further, the dark matter constraint and $\br(b \to s
\ga)$ would no 
longer be compatible for this value of $m_0$. We also see that, by
comparison with \ref{fig:hint1}(a), the $\br(B_s \to \mu^+\mu^-)$
constraint has moved to lower $A_0$, but the $\Mh$ constraint has dropped even
further, and $\Mh > 114 \gev$ over the entire visible plane. 
The net result is a region compatible with all the constraints
that extends from $A_0 \sim - 1850 \gev$ down to $A_0 \sim - 2150 \gev$
for $\mu \sim 400 \gev$, with a coannihilation filament extending to
larger $\mu$ when $A_0 \sim - 2200 \gev$. Once again all of the WMAP
strip in this panel is compatible with CDMS.

The larger allowed area of parameter space is reflected in panel (b)
of Fig.  \ref{fig:hint1}, which has $\mu = 400 \gev$ and $A_0 = - 2000 \gev$,
as well as $\tb = 45$ and $\MA = 160 \gev$ as before. In this
case, we see that a substantial region of the WMAP strip with 
$m_{1/2} \sim 700 \gev$ and a width $\delta m_{1/2} \sim 100 \gev$,
extending from $m_0 \sim 750 \gev$ to higher $m_0$ 
is allowed by all the other constraints. The
$\br(B_s \to \mu^+\mu^-)$ and $\Mh$ constraints have now moved to
relatively low values of $m_{1/2}$, but we still find 
$\br(B_s \to \mu^+\mu^-) > 2 \times 10^{-8}$ and $\Mh$ close to the
LEP lower limit. The $(g-2)_\mu$ constraint
truncates the allowed region at $m_0 \sim 1050 \gev$.  Once again,
the allowed region is compatible with CDMS (to the left of the (orange)
dotted line) provided the strangeness contribution to the proton mass is
small.  

Clearly our anticipation for the discovery of supersymmetry or new physics is great.
If the CDF hint of the Higgs boson is realized, we should expect to see numerous
departures from the standard model including the decay of $B_s$ to $\mu^+ \mu^-$,
$\br(b \to s \gamma)$ should show deviations from the standard model value,
and dark matter should be discovered by CDMS or XENON10 in the near future.
Similarly if an observation of new physics in any one of these areas occurs,
we should be ready for  flood of new data showing departures from the standard model.

\section*{Acknowledgments}
\noindent 
I would like to thank J. Ellis, T. Hahn, S. Heinemeyer, P. Sandick, Y. Santoso, 
V. Spanos, A. Weber, and G. Weiglein for the many collaborations whose results were
summarized here.
This work  was supported in part
by DOE grant DE--FG02--94ER--40823.

%
%

\end{document}
